\documentclass[12pt]{article}

\usepackage[english]{babel}
\usepackage[T1]{fontenc}

\usepackage[a4paper,top=2cm,bottom=2cm,left=2cm,right=2cm,marginparwidth=1.75cm]{geometry}

\usepackage{amsmath, amsthm, amssymb, nomencl, mathtools, algorithmic, csquotes, subcaption, float}
\usepackage[ruled]{algorithm2e}
\usepackage{graphicx}
\usepackage[natbib, maxcitenames=5, style=apa, uniquename=false, uniquelist=false]{biblatex}
\usepackage{appendix}
\usepackage{etoc}

\theoremstyle{plain}
\newtheorem{theorem}{Theorem}

\usepackage{setspace}

\theoremstyle{remark}
\newtheorem{remark}{Remark}

\usepackage[colorlinks=true, allcolors=blue]{hyperref}

\DeclareMathOperator*{\argmin}{arg\,min}

\DeclareMathOperator*{\cov}{cov}
\newcommand{\1}{\mathmybb{1}}
\allowdisplaybreaks

\DeclareMathAlphabet{\mathmybb}{U}{bbold}{m}{n}
\newcommand{\bX}{\boldsymbol{X}}
\newcommand{\bx}{\boldsymbol{x}}

\newcommand{\bh}{\boldsymbol{h}}
\newcommand{\bH}{\boldsymbol{H}}

\newcommand{\bz}{\boldsymbol{z}}

\newcommand{\bbeta}{\boldsymbol{\beta}}
\newcommand{\balpha}{\boldsymbol{\alpha}}

\newcommand{\logit}{\operatorname{logit}}

\makeatletter
\newcommand{\itemlabel}[1]{\def\@currentlabel{#1}}
\makeatother

\title{Data integration of non-probability and probability samples with deterministic predictive mean matching}

\author{Aniela Czerniawska\thanks{Adam Mickiewicz University in Pozna\'n; Poland, Statistical Office in Pozna\'n, Poland; Wieniawskiego 1, 61-712 Pozna\'n, Poland; ORCID: 0009-0001-5119-4891.} 
\and 
Piotr Chlebicki\thanks{Stockholm University, Sweden; Albano hus 1, 106 91 Stockholm, Sweden; ORCID: 0009-0006-4867-7434.} 
\and
{\L}ukasz Chrostowski\thanks{Analytic Partners, Krysiewicza 2, 61-887 Pozna\'n, Poland.}
\and 
Maciej Ber\k{e}sewicz\thanks{Pozna\'n University of Economics and Business; Department of Statistics, Al. Niepodleg{\l}o\'sci 10, 61-875 Pozna\'n, Poland, E-mail: \url{maciej.beresewicz@ue.poznan.pl}; Statistical Office in Pozna\'n, ul. Wojska Polskiego 27/29 60-624 Pozna\'n, Poland; ORCID: 0000-0002-8281-4301.}
}
\date{}

\begin{document}

\maketitle

\begin{center}
Running headline: PMM FOR DATA INTEGRATION
\end{center}

\section*{Acknowledgements}

\doublespacing

The authors' work has been financed by the National Science Centre in Poland, OPUS 20, grant no. 2020/39/B/HS4/00941. Codes to reproduce simulations from the paper are available at \url{https://github.com/ncn-foreigners/paper-nonprob-pmm}. Piotr Chlebicki developed the first version of the method as a~master's student; Aniela Czerniawska continued this work after Piotr graduated.

CRediT contributions:  Piotr Chlebicki -- Conceptualization, Methodology, Writing – original draft; Aniela Czerniawska -- Methodology, Validation, Writing – original draft, Writing – review \& editing; Łukasz Chrostowski: Software, Validation, Writing – original draft; Maciej Beręsewicz: Conceptualization, Funding acquisition, Methodology, Project administration, Supervision, Writing – original draft, Writing – review \& editing

\clearpage

\begin{abstract}
We study deterministic predictive mean matching mass imputation estimators to integrate data from probability and non-probability samples. We consider two approaches: predicted-to-predicted (PMM~A) and predicted-to-observed (PMM~B) matching. We prove the consistency of mean estimators, derive a variance decomposition, and propose estimators of variance. We establish consistency of the PMM~A estimator under model misspecification and underline key differences from the nearest neighbour method. Our PMM~B approach can be employed with non-parametric regression techniques, such as kernel regression, and the analytical expression for variance applies to nearest neighbour matching for non-probability samples. Extensive simulation studies compare properties of the proposed estimators with existing alternatives and examine the effects of model misspecification. The paper concludes with an empirical study on the integration of job vacancy survey and vacancies submitted to public employment offices (admin and online data). Open-source software is available.
\end{abstract}

Keywords: bootstrap, job vacancy survey, mass imputation, non-probability surveys, variance estimation
\clearpage

\doublespacing

\section{Introduction}

With the availability of large sets of administrative data, voluntary internet panels, social media and big data, inference with non-probability samples is being heavily studied in the statistical literature \citep{beaumont2020probability, elliott_inference_2017, berkesewicz2017two, citro2014multiple}. Because of their non-statistical character and unknown sampling mechanism, these sources cannot be used directly for estimating population characteristics. 

Several inference approaches have been proposed in the literature with respect to data from non-probability samples, which either involve data integration with population level data or probability samples from the same population \citep[for recent review see][]{wu2022statistical}. Main techniques for inference based on such data can broadly be classified into inverse probability weighting (IPW) estimators, prediction estimators (PE) and doubly robust (DR) estimators, which include both IPW and PE. In this paper we focus on mass imputation (MI) estimators, which are a variant of prediction estimators used for integrating probability and non-probability samples \citep[cf.][]{elliott_inference_2017}. The general idea is to impute values of the target variable $Y$ for all units in a probability sample based on values observed in non-probability samples (this is the $\xi p$ framework as discussed by \citet{wu2022statistical}). 

In recent papers two distinct MI estimators have been proposed: (1) the nearest neighbour (NN; cf. \citet{yang2021integration}) imputation estimator and (2) the parametric (PAR) and non-parametric (NPAR) imputation estimator \citep[cf.][]{kim_combining_2021,chen_nonparametric_2022}. The latter two consist of using a~model for $\mathbb{E}[Y|\bX]$ with parameters estimated from a non-probability sample and then using predictions from this model as imputed values for a~probability sample. In contrast, the NN imputation involves finding $k$ nearest neighbours for each unit in a probability sample from a set of units from a~non-probability sample (imputed values are observed values of $Y$ from the non-probability sample). 

In this paper we focus on the predictive mean matching (PMM) MI estimator in the context of data integration, which can be seen as a technique combining NN and parametric (or non-parametric) imputation estimators. PMM combined with multiple imputation was introduced by \citet{little1988missing,rubin1986statistical} for imputing missing data in surveys. \citet{schenker1996partial} and \citet{horton2001multiple} advocated PMM imputation because of its robustness to model misspecification. The asymptotic properties of the PMM estimator were recently discussed in the context of survey non-response by \citet{yang_asymptotic_2020}. Initial work on multiply robust PMM estimation was done by \citet{chen_note_2021}. However, the PMM estimator has not yet been discussed or studied in the context of data integration involving non-probability and probability samples. Our contribution can be summarised as follows: 

\begin{enumerate}
    \item We study properties of two variants of semi-parametric \emph{deterministic} PMM estimators for the data integration problem: matching either by predicted-predicted ($\hat{y}-\hat{y}$) or predicted-observed ($\hat{y}-y$) values. The first option is the standard PMM estimator, which is used in most of the existing studies, while the second one is less commonly discussed. The motivation for the $\hat{y}-y$ is based on its performance when continuous variables are used and the model is correctly specified. Moreover, we discuss mixed matching that involves both cases.
    \item We prove the consistency of these estimators under suitable assumptions where population size is either known or estimated. In the case of $\hat{y}-y$ estimator, the theorem can be extended to non-parametric models, such as Nadaraya–Watson kernel regression or even simple artificial neural networks. We provide proof of the consistency of the $\hat{y}-\hat{y}$ estimator under model misspecification. 
    \item We derive a variance decomposition and a partially analytic variance estimator and study its performance (along with fully bootstrap estimators) for finite populations. Furthermore, the proposed closed form expression for analytical variance can also be used for the NN estimator, extending the result of \citet{yang2021integration}.
\end{enumerate}

In the paper, we discuss the significant differences between the PMM estimator based on $\hat{y}-\hat{y}$ matching and NN estimator. The two are identical only in one very special case, when the \textit{assumed} model is linear and depends on a~single covariate. Of particular importance is the fact that predictive mean matching approach resolves the curse of dimensionality and uses less restrictive assumptions than the NN estimator. Our rigorous approach to study properties of the two proposed PMM estimators is different from those in the \citet{yang2021integration}. 

The structure of the paper is as follows. In Section~\ref{sec-basic} we provide basic notation, describe two PMM algorithms and discuss the assumptions underlying the PMM estimators. Section~\ref{sec-main} provides proof of consistency of the PMM estimators and the exact variance expression along with its estimators. Section \ref{sec-sim} describes one of the simulation studies, whose results verify the proposed approach under simple random sampling where two linear and non-linear models are considered. Section~\ref{sec-empirical} presents our empirical study. The article ends with conclusions and the Appendix including details concerning the proofs as well as additional simulation studies. All calculations in the article can be reproduced using R \citep{r-cran} and the \texttt{nonprobsvy} package \citep{nonprobsy-pkg, nonprobsvy-paper} developed by the authors. All codes are available at \url{https://github.com/ncn-foreigners/paper-nonprob-pmm}.

\section{Basic setup}\label{sec-basic}

\subsection{Notation and motivation}

Let $U=\{1,..., N\}$ denote the target population consisting of $N$ labelled units. Each unit $i$ has an associated vector of auxiliary variables $\bx_{i}$ (a realisation of the random vector $\bX_{i}$ in the superpopulation) and the study variable $y_{i}$ (a realisation of the random variable $Y_{i}$ in the superpopulation). Let $\{ (y_i, \bx_i), i \in S_{\text{NP}}\}$ be a dataset of a non-probability sample of size $n_{\text{NP}}$ and let  $\{\left(\bx_i, \pi_{i}\right), i \in S_{\text{P}}\}$ be a dataset of a probability sample of size $n_{\text{P}}$, where only information about variables $\bX$ and inclusion probabilities $\pi$ (which in the superpopulation model are also considered to be random variables) is available. Let $\delta$ be an indicator of inclusion into non-probability sample. Each unit in the sample $S_{\text{P}}$ has been assigned a~design-based weight given by $d_i = 1/\pi_i$. The setting is summarised in Table \ref{tab-two-sources}. 

\begin{center}
    Table \ref{tab-two-sources} around here.
\end{center}

Finally, throughout the entire paper, we use the symbols $\mathbb{P}, \mathbb{E}, \text{cov}$ to refer to the probability measure, expectation and covariance operators respectively with respect to $\mathbb{P}$, where the randomness comes from both the superpopulation model $Y, X, \pi$ and the finite population $\delta, \1_{i\in S_{\text{NP}}}|\pi$.
We also use the standard notation for compound random variable created from a~set of rvs $\{Z_{i}; i\}$ in index space $W$ as:
\begin{equation*}
    Z_{W}:=Z_{W(\omega)}(\omega)=\begin{cases}
        Z_{1}(\omega) & \text{when } W(\omega) = 1\\
        Z_{2}(\omega) & \text{when } W(\omega) = 2\\
        \vdots
    \end{cases}
\end{equation*}

The goal is to estimate a~finite population mean $\mu_{y}=\frac{1}{N}\sum_{i=1}^{N} y_{i}$ of the target variable $Y$. As values of $y_{i}$ are not observed in the probability sample, it cannot be used to estimate the target quantity. Instead, one could try combining the non-probability and probability samples to estimate $\mu_{y}$. In this paper, we do not consider modifications for the possible overlap.

In our paper, we focus on prediction, which is an appealing solution as it allows us to make micro-level data available with fully imputed target variables and design specification. Moreover, we impute \textit{observed} values of $Y$, which makes it easier to explain them to users of the micro-data (since observations are not generated from an \textit{unknown} model).

\subsection{Predictive mean matching algorithms}

In this section, we describe two general algorithms for semi-parametric PMM estimators. Let $d$ denote a bivariate distance function on $\mathbb{R}$ (usually a~semi-metric). The estimators we are interested in are a result of the following algorithm:

\begin{algorithm}[ht!]
\small
\caption{Two deterministic predictive mean matching imputation algorithms}
\label{algo-1}\DontPrintSemicolon
\nlset{1} Estimate $\bbeta$ in regression model $\mathbb{E}[Y|\bX=\bx]=m(\bx, \bbeta)$, we denote this estimate by $\hat{\bbeta}$.\;

\nlset{2A} ($\hat{y}-\hat{y}$ imputation; PMM A) Impute $\hat{y}_{i}=m\left(\bx_{i},\hat{\bbeta}\right), 
\hat{y}_{j}=m\left(\bx_{j},\hat{\bbeta}\right)$ 
for $i\in S_{\text{P}}, j\in S_{\text{NP}}$ and map each 
$i\in S_{\text{P}}$ to $\hat{\nu}(i)$ where
$\hat{\nu}(i)=
\argmin_{j\in S_{\text{NP}}}d\left(\hat{y}_{i},\hat{y}_{j}\right)$. 

If $k>1$, let:
\begin{equation}\label{y-nu-definition-yhat-yhat}
    \hat{\nu}(i, z) = \argmin_{\displaystyle j\in S_{\text{NP}}\setminus\bigcup_{t=1}^{z-1}
    \{\hat{\nu}(i, t)\}} d\left(\hat{y}_{i},\hat{y}_{j}\right),
\end{equation}
i.e. the $\hat{\nu}(i, z)$ is the $z$'th nearest neighbour from the sample.\;

\nlset{2B} ($\hat{y}-y$ imputation; PMM B) Impute $\hat{y}_{i}=m\left(\bx_{i},\hat{\bbeta}\right)$ 
for $i\in S_{\text{P}}$ and map each 
$i\in S_{\text{P}}$ to $\hat{\nu}(i)$ where $ \hat{\nu}(i)=
\argmin_{j\in S_{\text{NP}}}d\left(\hat{y}_{i},y_{j}\right)$. 

If $k>1$, let:
\begin{equation}\label{y-nu-definition-yhat-y}
    \hat{\nu}(i, z) = \argmin_{\displaystyle j\in S_{\text{NP}}\setminus\bigcup_{t=1}^{z-1}
    \{\hat{\nu}(i, t)\}}
    d\left(\hat{y}_{i},y_{j}\right).
\end{equation}\;

\nlset{3} Use the imputed values in estimation: 
$ \hat{\mu}=
N^{-1}\sum_{i \in S_{\text{P}}}\pi_{i}^{-1} y_{\hat{\nu}(i)}$, or if $k>1$, then:
\begin{equation}\label{mu-hat-definition-yhat-yhat-match}
    \hat{\mu}=\frac{1}{N}\sum_{i \in S_{\text{P}}}
    \frac{1}{\pi_{i}}\frac{1}{k}\sum_{t=1}^{k}y_{\hat{\nu}(i, t)}.
\end{equation}\;
\end{algorithm}

We denote the $k$ nearest neighbours for unit $i$ by $\hat{\nu}(i, t), t=1\ldots,k$ to emphasise that it is a random variable with values in the index space. If the nearest neighbours are not unique, then the ties are broken by random selection.

Algorithm \ref{algo-1} (without step 2B) represents the standard $\hat{y}-\hat{y}$ matching. Throughout the paper, this estimator is denoted as the $\hat{y}-\hat{y}$ or as the PMM~A estimator, and the resulting estimator is given by the expression \eqref{mu-hat-definition-yhat-yhat-match}. 

Algorithm \ref{algo-1} (without step 2A) is a~$\hat{y}-y$ matching imputation estimator, in which units from the probability sample $S_{\text{P}}$ are matched based on $\hat{y}$ to those from the non-probability sample $S_{\text{NP}}$ based on $y$. This estimator is denoted as the $\hat{y}-y$ or the PMM B estimator. The resulting estimator is given by the same expression \eqref{mu-hat-definition-yhat-yhat-match}, but the matching is done differently.

Note that in the PMM A algorithm, we ensure that the $\hat{y}$ value is similar for the matched units. The residuals do not affect the donor selection, and only in the next step are they averaged across the selected donors. Therefore, their contribution depends on how the donor weights are distributed. 

On the other hand, in the case of PMM B, the matching being done on $y = \hat{y} + \varepsilon$ implies that the residual component directly influences the choice of donors. In this setting, the donors may not be selected only because the regression function values are close, but also because of the residual, which can compensate for the differences in $m(\bX, \bbeta)$. This implies some additional variability that might not vanish asymptotically with the fixed number of donors $k$. These differences are reflected in the assumptions and are discussed in the following sections.

\subsection{Comparison with the nearest neighbour imputation approach}

If we let $m(\bx_{i}, \bbeta)=m(x_{i}, \beta)=x_{i}\beta$, i.e. use only one covariate, the resulting estimator \eqref{mu-hat-definition-yhat-yhat-match} for $\hat{y}-\hat{y}$ matching is exactly the nearest neighbour estimator, since linear scaling preserves distances on $\mathbb{R}_{+}$. This is the only case in which these two coincide. When $m$ is linear regression and $d$ is a~Euclidean distance, the $\hat{y}-\hat{y}$ matching could be thought of as a~"pseudo-nearest neighbour" matching with a weighted (by estimated regression coefficients) distance function $d':(\bz_{i},\bz_{j})\mapsto\left|(\bz_{i}-\bz_{j})^{T}\hat{\bbeta}\right|$ which expands to:
\begin{equation*}
    d'(\bz_{i}, \bz_{j})=\left|(\bz_{i}-\bz_{j})^{T}\left(\sum_{t\in S_{\text{NP}}}\bX_{t}\bX_{t}^{T}\right)^{-1}
    \sum_{t\in S_{\text{NP}}}\bX_{t}Y_{t}\right|,
\end{equation*}
where $\mathbb{X}$ denotes the matrix of covariates in $S_{\text{NP}}$, and $\boldsymbol{Y}$ denotes the vector of response variables in $S_{\text{NP}}$. Since $d'$ depends on $y$ values from $S_{\text{NP}}$, one cannot simply assert that the $\hat{y}-\hat{y}$ matching estimator inherits properties from the NN estimator, which requires the distance function to be deterministic. Similarly, in the general case of GLM regression with link function $g$, the distance -- not necessarily a metric -- is given by:
\begin{equation*}
    d'(\bz_{i}, \bz_{j})=\left|g^{-1}\left(\bz_{i}^{T}\hat{\bbeta}\right)-g^{-1}\left(\bz_{j}^{T}\hat{\bbeta}\right)\right|,
\end{equation*}
with $\hat{\bbeta}$ depending on the values of $(Y, \bX)$ from $S_{\text{NP}}$. In contrast, the PMM B estimator based on the $\hat{y}-y$ matching does not coincide with NN in any widely recognizable general setting.

In general, despite the similarity of PMM and NN in the description of the algorithm, there are more differences than similarities between the properties of their corresponding imputations. One key difference is that PMM does not suffer from the curse of dimensionality since our assumptions do not place many restrictions on the number of covariates. In fact, assumptions for $\hat{y}-\hat{y}$ matching are more readily attainable with a higher number of covariates (cf. \citet{yang_asymptotic_2020}). Moreover, we do not require positivity, contrary to the NN estimator, and later on, we show in simulation studies in Appendix \ref{sec-appen-sim-4} that the PMM estimators work well under stochastic and deterministic under-coverage (with one exception).

\begin{remark}(Regarding $\hat{y}-y$ matching)
\label{remark-1}
    If the study variable $Y$ is a binary variable, then the $\hat{y}-y$ matching estimator \eqref{mu-hat-definition-yhat-yhat-match} is just a modification of the GLM estimator, in which, instead of predicted values, we impute a classification for each $i\in S_{\text{P}}$. The resulting estimate may still be consistent according to Theorem~\ref{yhat-y-consistency-proof}, but it may have comparatively poor finite sample properties in practice. 
\end{remark}

\subsection{Assumptions}\label{sec-assumptions}

We state the main assumptions used for studying the properties of estimators defined in \eqref{mu-hat-definition-yhat-yhat-match}, with the distance function being the Euclidean metric. See \citet{kim_combining_2021, yang_asymptotic_2020} for a comparison with the assumptions for other mass imputation estimators.

Throughout the paper, we assume that in the superpopulation model every $Y$ variable has a finite variance, i.e. $\forall i:\mathbb{E}\left[Y_{i}^{2}\right]<\infty$.
\begin{itemize}
    \item[(A1)] \itemlabel{A1}\label{conditional-mean-assum} $\mathbb{E}[Y_{i}|\bX_{i}]=m\left(\bX_{i},\bbeta_{0}\right)$  almost surely for some $\bbeta_{0}\in\mathbb{R}^{p}$ and continuous $m$ (as a functional from $\mathbb{R}^{p}\times\mathbb{R}^{p}$), with $\hat{\bbeta}$ being a consistent estimator for $\bbeta_{0}$.
    \item[(A2)] \itemlabel{A2}\label{lim-res-prob-assum}
    We require that for the true vector of regression coefficients $\bbeta_{0}$, we have:
    \begin{equation*}
        \lim_{N\rightarrow\infty}
        \frac{1}{N}\sum_{i=1}^{N}\left(m(\bX_{i},\bbeta_{0})-Y_{i}\right)=0
        \text{ in probability.}
    \end{equation*}
    \item[(A3)] \itemlabel{A3}\label{lipschitz-assum} We assume that there exists a neighbourhood $\mathcal U$ of $\beta_0$ such that:
    \begin{equation*}
        |m(\bX, \bbeta) - m(\bX, \bbeta')| \leq L(\bX) \| \bbeta - \bbeta'\|, \quad \forall \bbeta, \bbeta' \in \mathcal U,
    \end{equation*}
    where $L(\bX)$ is an integrable bound and $E[L(\bX)] < \infty$.
    \item[(A4)] \itemlabel{A4}\label{asymptotic-neighbour-assum}
    Let $M_{j, i_{1}, \dots, i_{k}}$ denote an event measurable with respect to the $\sigma$-algebra generated by $\{\1_{\{t \in S_{\text{NP}}\}}, \1_{\{t \in S_{\text{P}}\}}, \bx_t, y_t : t \in {j, i_1, i_2, \dots}\}$. The event $M_{j, i_{1}, \dots, i_{k}}$ depends only on units $j, i_{1}, \dots, i_{k}$ in the population. We require that whenever $\mathbb{P}(M_{j, i_{1}, \dots, i_{k}})>0$ for infinitely many tuples $(i_{1}, \dots, i_{k})$ and some unit $j$ from the superpopulation, then:
    \begin{equation*}
        \lim_{n_{\text{NP}}\rightarrow\infty}\mathbb{P}[1\leq \#\{i_{1}, \dots, i_{k}\in S_{\text{NP}}, j: M_{j, i_{1}, \dots, i_{k}}\}|\{j\in S_{\text{P}}\}]=1.
    \end{equation*}  
    \item[(A5)] \itemlabel{A5}\label{pop-sequence-assum} We~study the asymptotic properties of \eqref{mu-hat-definition-yhat-yhat-match} in the setting where we consider a sequence of populations such that:
    \begin{equation*}
        \exists C_{1}, C_{2} \forall i: C_{1}\leq\frac{N\pi_{i}}{n_{\text{P}}}\leq C_{2}\text{ almost surely}.
    \end{equation*}
    Each population in the sequence is a random sample from the superpopulation and $\displaystyle\lim_{n_{\text{NP}}\rightarrow\infty}n_{\text{P}}=\infty$ a.s.
    \item[(B1)] \itemlabel{B1}\label{yhat-yhat-assum} (For $\hat{y}-\hat{y}$ matching only) For any $\varepsilon>0$, we assume that:
    \begin{equation*}
        \forall i:\lim_{n_{\text{NP}}\rightarrow\infty}
        \mathbb{P}\left[\left.\exists j_{1},j_{2},\dots, j_{k}\in S_{\text{NP}}:
        \max_{t}\lVert\bX_{i}-\bX_{j_{t}}\rVert<\varepsilon\right|
        i\in S_{\text{P}}\right]= 1,
    \end{equation*}
    i.e. asymptotically we will cover at least $k$ units that have similar $\bX$ values to $i\in S_{\text{P}}$.
    \item[(B2)] \itemlabel{B2}\label{linear-rep-assum} (For $\hat{y}-\hat{y}$ matching only) The estimator $\hat{\bbeta}$ has a linear representation of the form:
    \begin{equation*}
        \hat\bbeta - \bbeta_0 = \frac{1}{n_{\text{NP}}} \sum_{j \in S_{\text{NP}}} \psi(Y_j, \bX_j) + o_P(n_{\text{NP}}^{-1/2})
    \end{equation*}
    where the influence function $\psi(\cdot)$ satisfies $\mathbb{E}[\psi(Y, \bX)] = 0$ and $\mathbb{E}[\|\psi(Y, \bX)\|^2]$ exists.
    \item[(B3)] \itemlabel{B3}\label{donor-reuse-assum} (For $\hat{y}-\hat{y}$ matching only) Let
    \begin{equation*}
        W_{j,N} = \frac{1}{N} \sum_{i \in S_{\text{P}}} \frac{1}{\pi_i} \frac{1}{k} \sum_{t=1}^k \1 \{\hat{\nu}(i,t) = j\}, \quad j \in S_{\text{NP}}
    \end{equation*}
    denote the fraction of the overall donor weight contributed by unit $j$. We assume that: 
    \begin{equation*}
        \sum_{j \in S_{\text{NP}}} W_{j,N}^2 \overset{P}{\rightarrow} 0.
    \end{equation*}
    \item[(C1)] \itemlabel{C1}\label{yhat-y-assum} (For $\hat{y}-y$ matching only) The support of $\bX$ in the superpopulation is such that for almost all $i$ and all $\varepsilon > 0, y$:
    \begin{equation*}
        \mathbb{P}\left[Y_{i}\in
        \left(y-\varepsilon, y+\varepsilon\right)\right]>0\implies
        \mathbb{P}\left[m(\bX_{i}, \bbeta)\in
        \left(y-\varepsilon, y+\varepsilon\right)\right]>0,
    \end{equation*}
    for $\bbeta$ in some arbitrarily small open neighbourhood of $\bbeta_{0}$.
\end{itemize}

\subsection{Discussion regarding the assumptions}

Although we do not impose the missing at random (MAR) assumption,  the existence of a~consistent estimator $\hat\bbeta$ in \eqref{conditional-mean-assum} requires additional conditions. One commonly used sufficient condition is $\mathbb{P}[Y\in dy|\bX,\delta]=\mathbb{P}[Y\in dy|\bX]$, which is essentially the MAR assumption. This condition and other assumptions that are sufficient for the consistency of the regression estimator $\hat{\bbeta}$ in (\ref{conditional-mean-assum}) are discussed in more detail in Appendix \ref{app-consistency-reg}. Importantly, MAR is not required for the results presented in this paper, as long as a consistent $\hat\bbeta$ exists. Later, we relax the assumption (\ref{conditional-mean-assum}) while discussing the consistency of PMM estimators under model misspecification.

We also do not assume positivity. Instead, a weaker set of assumptions: \eqref{asymptotic-neighbour-assum} along with  \eqref{yhat-yhat-assum} or \eqref{yhat-y-assum} (depending on the PMM estimator type) serves as a replacement for the standard positivity condition. The interpretation of \eqref{asymptotic-neighbour-assum} and the discussion about the positivity assumption are presented in Section \ref{sec-assum-nonprob}.

Assumption (\ref{pop-sequence-assum}) describes a setting in which we explore consistency and is not important for finite sample performance.

\subsubsection{Assumptions on the outcome model and covariates}\label{sec-assum-covariates}

Assumption \eqref{lim-res-prob-assum} ensures that asymptotically, the average of the regression residuals disappears, meaning that the regression model correctly captures the population mean of $Y$. In the proof strategy, we use \eqref{lim-res-prob-assum} to first show the consistency of the mass imputation estimator of the population mean. This result is then used as a foundation for showing the consistency of both PMM estimators. Note that this assumption is not implied by \eqref{conditional-mean-assum}, as \eqref{conditional-mean-assum} restricts only the conditional mean of $Y$ given $\bX$ but does not address the dependence structure of the residuals $\varepsilon_i = Y_i - m(\bX_i, \bbeta_0)$.

Assumption \eqref{lipschitz-assum} imposes a local Lipschitz condition on the regression function $m(\bX, \bbeta)$ around the true parameter $\bbeta_0$. Intuitively it means that a small deviation of $\hat\bbeta$ from $\bbeta_0$ leads to a small change in the predicted values. In our case, this condition ensures that the donor selection and resulting imputations are stable.

Assumptions (\ref{yhat-yhat-assum}) and (\ref{yhat-y-assum}) are not strictly necessary for reasonable performance in finite samples, but the PMM estimator will perform better if the non-probability sample is reasonably diverse, which is what assumptions (\ref{yhat-yhat-assum}) and (\ref{yhat-y-assum}) are essentially implying. Assumption (\ref{yhat-y-assum}) is satisfied if, for example, for almost all units $i$, one covariate $x_{j}$ has support on the whole real line, $m(\cdot,\bbeta)$ depends on this covariate, and the closure of the range of function $x_{j}\mapsto m\left(\bx,\bbeta\right)$ contains the support of $Y$. For instance, in the case of logistic regression, where the closure of the range of:
\begin{equation*}
    x_{j}\mapsto(1+\exp(-\beta_{0}-\beta_{1}x_{1}-\ldots-\beta_{j}x_{j}-\ldots-\beta_{p}x_{p}))^{-1},
\end{equation*}
is $[0,1]\supseteq\{0, 1\}$. This condition is stated for almost all units, since for some values of $\bX$, the implication may not be true. For example, in the case of a linear regression model without intercept, if $\bX =0$, then it is not guaranteed that $\mathbb{P}(m(0, \bbeta) \in (y -\varepsilon, y + \varepsilon)) > 0$ for all $\bbeta$ unless $0 \in (y -\varepsilon, y + \varepsilon)$.

Assumption \eqref{linear-rep-assum} ensures that the influence of any observation in $S_{\text{NP}}$ on $\hat\bbeta$ is small. This assumption is especially relevant for the PMM~A method, because $\hat\bbeta$ is estimated from the same sample $S_{\text{NP}}$ from which the donors are selected, and it introduces a dependence between the donor selection and the estimation error in $\hat\bbeta$ (meaning that the residual $\varepsilon_j = Y_j - m(\bX_j, \bbeta_0)$ of a potential donor and the estimated $\hat\bbeta$ are not necessarily independent). Assumption \eqref{linear-rep-assum} ensures that this dependence is negligible asymptotically. Note that this problem does not arise for the PMM~B as in that algorithm the donors are selected based on the observed $Y_j$, so the residuals and $\hat\bbeta$ are independent by construction of the method. Further discussion and sufficient conditions are presented in the Appendix \ref{app-linear-rep}.

\subsubsection{Assumptions on non-probability sampling}\label{sec-assum-nonprob}

In proving consistency, we do not assume positivity, contrary to \citet{kim_combining_2021} and \citet{yang2021integration}. Instead, a very mild regularity condition \eqref{asymptotic-neighbour-assum} regarding the quality of non-probability sampling is required, along with less restrictive assumptions \eqref{yhat-yhat-assum}/\eqref{yhat-y-assum} depending on the PMM estimator.

Assumption (\ref{asymptotic-neighbour-assum}) can be understood as follows. Suppose that unit $j$ has the property that, for infinitely many tuples of units in the superpopulation $(i_{1}, \ldots, i_{k})$, some condition, for example $\left|Y_{j}-\frac{1}{k}\sum_{t=1}^{k} Y_{i_{t}}\right|<\varepsilon$, is satisfied with positive probability. Then, as the sample size of $S_{\text{NP}}$ tends to $\infty$, we will almost surely (conditional on $j\in S_{\text{P}}$) be able to find at least one such tuple $(i_{1}, \ldots, i_{k})$, where $i_{t}\in S_{\text{NP}}$ for each $t \in \{1, \dots, k\}$, that satisfies the condition. This assumption is used to ensure that, in conjunction with other assumptions, finding a~set of neighbours in the non-probability sample that represent units in the probability sample is asymptotically possible.

In the case of discrete $\bX$ (such as age or gender), we do not require that for all units $\displaystyle\mathbb{P}[\delta=1|\bX = \bx_{i}]>0$, as may happen in quota sampling or Internet surveys or via dependence on some unobserved variables, where certain units may be missing due to a lack of access to the Internet. Instead, we only require that each level of $\bX$ will \textit{eventually} be represented by at least $k$ units as the sample size tends to infinity. 

The case of continuous variables is analogous, i.e. if we set $i\in U$ and $\varepsilon>0$, we require that eventually sample $S_{\text{NP}}$ will contain at least $k$ units such that the distance between $\bx_{i}$ and $\bx_{j}$ corresponding to these $k$ units will be no greater than $\varepsilon$.

\subsubsection{Assumptions on donor usage}\label{sec-assum-donors}
Assumption \eqref{donor-reuse-assum}, required only for PMM A matching, ensures that no single donor dominates. In the setting of PMM A, donors are selected based on their similarity to the predicted values rather than the observed outcomes (PMM B). Therefore, the residuals, $\varepsilon_j = Y_j - m(\bX_j, \bbeta_0)$, are not controlled in PMM A. Consequently, the estimator can be decomposed as:
\begin{equation*}
    \frac{1}{N} \sum_{i \in S_{\text{P}}} \frac{1}{\pi_i} \frac{1}{k} \sum_{t=1}^k Y_{\hat{\nu}(i,t)} = \sum_{j \in S_{\text{NP}}} W_{j,N} Y_j = \sum_{j \in S_{\text{NP}}} W_{j,N} m(\bX_j, \bbeta_0) +\sum_{j \in S_{\text{NP}}} W_{j,N} \varepsilon_j,
\end{equation*}
where $W_{j,N}$ is the donor weight as defined in assumption \eqref{donor-reuse-assum}. Without some additional control, these residuals can accumulate if a few donors are repeatedly selected, potentially preventing consistency. We assume \eqref{donor-reuse-assum} as a remedy to this problem. The intuition is that while the matching mechanism in PMM A controls the predicted values $m(\bX_j, \hat\bbeta_j)$, as $n_{NP}$ grows, the residuals $\varepsilon_j$ are spread across sufficiently many distinct donors to vanish asymptotically.

\section{Main results}\label{sec-main}

\subsection{Consistency under correctly specified model}\label{sec-main-correct-spec}

The proof strategy, which is a~standard strategy for non-probability samples, can essentially be reduced to first proving that \eqref{mu-hat-definition-yhat-yhat-match} and the Horvitz-Thompson (HT) estimator (which we cannot construct) have the same limit, and then inferring the consistency of \eqref{mu-hat-definition-yhat-yhat-match} from the consistency of the \textit{hypothetical} HT in the setting described in Section \ref{sec-basic}.

\begin{theorem}\label{yhat-y-consistency-proof}
    If (\ref{conditional-mean-assum}), (\ref{lipschitz-assum}), (\ref{asymptotic-neighbour-assum}), (\ref{pop-sequence-assum}), (\ref{yhat-y-assum}) hold and the model is correctly specified, then the mass imputation estimator:
    \begin{equation*}
        \hat{\mu}_{\text{MI}}= \frac{1}{N}\sum_{i\in S_{\text{P}}}\frac{1}{\pi_{i}}m\left(\bX_{i}, \hat{\bbeta}\right)
    \end{equation*}
    is consistent if and only if the $\hat{y}-y$ matching estimator in \eqref{mu-hat-definition-yhat-yhat-match} is consistent. Moreover, assumptions (\ref{conditional-mean-assum}), (\ref{lim-res-prob-assum}), (\ref{lipschitz-assum}) are sufficient for the consistency of $\hat{\mu}_{\text{MI}}$.
\end{theorem}

\begin{theorem}\label{yhat-yhat-consistency-proof}
Under correctly specified regression model $m(\cdot, \bbeta)$ and assumptions (\ref{conditional-mean-assum}), (\ref{lipschitz-assum}), (\ref{asymptotic-neighbour-assum}), (\ref{pop-sequence-assum}), (\ref{yhat-yhat-assum}), (\ref{linear-rep-assum}), (\ref{donor-reuse-assum}), the $\hat{y}-\hat{y}$ estimator in \eqref{mu-hat-definition-yhat-yhat-match} is consistent when $\hat{\mu}_{\text{MI}}$ is consistent. Assumptions (\ref{conditional-mean-assum}), (\ref{lim-res-prob-assum}), (\ref{lipschitz-assum}) are sufficient for the consistency of $\hat{\mu}_{\text{MI}}$.
\end{theorem}

Proofs of Theorems \ref{yhat-y-consistency-proof}, \ref{yhat-yhat-consistency-proof} are given in Appendix \ref{proof-of-yhat-y} and \ref{proof-of-yhat-yhat} respectively. Theorem \ref{yhat-yhat-consistency-proof} does not address the necessity of consistency of $\hat{\mu}_{\text{MI}}$, unlike Theorem \ref{yhat-y-consistency-proof}. Additionally, if $N$ is unknown and estimated by $\sum_{r\in S_{\text{P}}}\pi_{r}^{-1}$ (i.e. H\'{a}jek estimator), the corrected estimates for $\hat{\mu}$ from \eqref{mu-hat-definition-yhat-yhat-match} are also consistent.

Since $\hat{\mu}$ and the \textit{hypothetical} HT estimator have the same limit, in order for \eqref{mu-hat-definition-yhat-yhat-match} to be asymptotically normal with $\left(\mu^{(HT)}_{n_{\text{P}}}, V^{(HT)}_{n_{\text{P}}}\right)$, it is both sufficient and necessary for the HT estimator to be asymptotically normal with $\left(\mu^{(HT)}_{n_{\text{P}}}, V^{(HT)}_{n_{\text{P}}}\right)$.

\begin{remark}(Regarding assumption (\ref{conditional-mean-assum}))
\label{remark-2}
 We may relax assumption (\ref{conditional-mean-assum}) by assuming that for continuous $m$: $\mathbb{E}[Y|\bX]=m(\bX)$ almost surely, i.e. a continuous function is a parameter, not just a vector in the Euclidean space. Then, under modified assumption (\ref{conditional-mean-assum}): $\hat{m}$ is consistent for $m$ in the sense that:
\begin{equation*}
        \forall\varepsilon>0:
        \lim_{n_{\text{NP}}\rightarrow\infty}
        \mathbb{P}\left[\lVert m-\hat{m}\rVert_{\infty}<\varepsilon\right]=
        \lim_{n_{\text{NP}}\rightarrow\infty}
        \mathbb{P}\left[\sup_{\bx\in \mathbb{R}^{p}}|m(\bx)-\hat{m}(\bx)|<\varepsilon\right]=1,
\end{equation*}
and almost surely continuous, we can establish the consistency of \eqref{mu-hat-definition-yhat-yhat-match} for PMM~B, where matching is done using $\hat{m}$ with almost exactly the same argument. This means that certain non-parametric regression techniques, such as kernel regression via the Nadaraya–Watson estimator, local linear regression techniques or even neural networks, could potentially replace the parametric regression step in PMM~B under suitable assumptions, see e.g. \citet{tyrcha_mielniczuk} for discussion of (stronger type) consistency for perceptrons. In the Appendix~\ref{sec-appen-sim-3}, we present a simulation study that covers some examples.
    
For PMM~A, assumption \eqref{linear-rep-assum} does not transfer directly to the non-parametric estimators, but an analogue could be a leave-one-out stability condition: $\|\hat{m} - \hat{m}^{(-j)}\| = O_P(n_{\text{NP}}^{-1})$. However, verifying this for specific non-parametric methods may be difficult in practice. The simulation result for PMM~A in the Appendix~\ref{sec-appen-sim-3} should therefore be regarded as exploratory rather than formal confirmation of the theoretical properties.
    
A question of whether convergence in $\lVert\cdot\rVert_{\infty}$ can be replaced by convergence in a weaker topology could be subject to further research.
\end{remark}

\subsection{Consistency under misspecified model}\label{sec-main-misspecified}

\citet{yang_asymptotic_2020} discussed the robustness of the PMM estimator for the imputation of item non-response in survey sampling, showing consistency under model misspecification when the support for $m(\bX, \bbeta_0)$ with fixed $\bbeta$ is convex and compact and the missing at random assumption holds. The following Theorem provides conditions under which the PMM~A estimator for data integration is consistent under misspecification (the proof is given in the Appendix).

\begin{theorem}\label{yhat-yhat-consistency-mis-specified}
    Let $\mathbb{E}\left[Y|\bX\right]=m'(\bX, \bbeta_{0})$ and the $\hat{y}-\hat{y}$ matching is done via the working model $m$ where possibly $m'\not\equiv m$. Assume that (\ref{lim-res-prob-assum}) holds for $m'$, (\ref{asymptotic-neighbour-assum}), (\ref{pop-sequence-assum}), (\ref{yhat-yhat-assum}), (\ref{linear-rep-assum}), (\ref{donor-reuse-assum}) hold along with:
    \begin{itemize}
        \item[(D1)] \itemlabel{D1}\label{miss-limit-assum} $\displaystyle\bbeta^{\ast}=\lim\hat{\bbeta}$ in probability exists,
        \item[(D2)] \itemlabel{D2}\label{miss-exp-assum}   $\displaystyle \mathbb{E}\left[ \left|Y_{i}-Y_{j}\right| \mid m(\bX_{i}, \bbeta^{\ast}), m(\bX_{j}, \bbeta^{\ast})\right]\leq
        C\left|m(\bX_{i}, \bbeta^{\ast})-m(\bX_{j}, \bbeta^{\ast})\right|$ almost surely.
    \end{itemize}
    Furthermore, assume that (\ref{lipschitz-assum}) holds on a neighbourhood $\mathcal{U}$ containing both $\bbeta_0$ and $\bbeta^*$. Then the $\hat{y}-\hat{y}$ matching estimator is consistent. Additionally, if the MI estimator based on model $m'$ converges in $L^{1}$ norm, then the $\hat{y}-\hat{y}$ estimator also converges in $L^{1}$ norm.
\end{theorem}

\begin{remark}(Regarding the Theorems \ref{yhat-y-consistency-proof}, \ref{yhat-yhat-consistency-proof}, \ref{yhat-yhat-consistency-mis-specified})
\label{remark-3}
    If $Y$ is multivariate we can, under analogous assumptions (though these assumptions may be somewhat less realistic in a multivariate case), replace $|\cdot|$ norm by the Euclidean norm $\lVert\cdot\rVert$ and get the consistency of the PMM estimators in a multivariate case.
    
    Additionally, under reasonable restrictions the proofs can be rewritten to accommodate distance functions other than the Euclidean metric almost in an exactly analogous manner.
\end{remark}

\subsection{Variance}

Let us consider the variance of \eqref{mu-hat-definition-yhat-yhat-match}, focusing on the case when $Y$ is univariate for simplicity. The results can easily be generalised to multivariate $Y$ by replacing products of scalars with outer products of vectors.

\begin{theorem}\label{variance-pmm-proof}
If the population size $N$ is known, the variance of estimators \eqref{mu-hat-definition-yhat-yhat-match} is given by
\begin{equation}
    \mathbb{V}\left[\hat{\mu}\right]=V_1 + V_2,
    \label{eq-var-est-theorem}
\end{equation}
\noindent where 
\begin{align}\label{V-1-expression}
    V_1&=
    \frac{1}{N^{2}k^{2}}
    \sum_{i=1}^{N}
    \mathbb{E}\left[
    \left(\frac{1}{\pi_{i}}-1\right)
    \left(\sum_{t=1}^{k}Y_{\hat{\nu}(i, t)}\right)^{2}\right]\nonumber\\
    &+\frac{1}{N^{2}k^{2}}\sum_{\substack{i,j=1\\i\neq j}}^{N}
    \mathbb{E}\left[\pi_{i}^{-1}\pi_{j}^{-1}
    \cov (\1_{\{i\in S_{\text{P}}\}},\1_{\{j\in S_{\text{P}}\}})
    \sum_{t',t=1}^{k}Y_{\hat{\nu}(i, t)}Y_{\hat{\nu}(j, t')}\right]\nonumber\\
    &=
    \mathbb{E}\left[
    \frac{1}{N^{2}}\sum_{i=1}^{N}\sum_{j=1}^{N}
    \frac{\pi_{ij}-\pi_{i}\pi_{j}}{\pi_{i}\pi_{j}}
    \left(\frac{1}{k}\sum_{t =1}^{k}Y_{\hat{\nu}(i, t )}\right)
    \left(\frac{1}{k}\sum_{t'=1}^{k}Y_{\hat{\nu}(j, t')}\right)\right],
\end{align}
\noindent and 
\begin{equation}\label{V-2-expression}
    V_2=\frac{1}{N^{2}}
    \sum_{i=1}^{N}\sum_{j=1}^{N}
    \cov\left(
    \frac{1}{k}\sum_{t =1}^{k}Y_{\hat{\nu}(i, t)},
    \frac{1}{k}\sum_{t'=1}^{k}Y_{\hat{\nu}(j, t')}\right)=
    \mathbb{V}\left[\frac{1}{N}
    \sum_{i=1}^{N}
    \frac{1}{k}\sum_{t=1}^{k}Y_{\hat{\nu}(i, t)}
    \right]. 
\end{equation}
\end{theorem}
The proof of Theorem \ref{variance-pmm-proof} is given in Appendix \ref{proof-of-2}. The $V_{1}$ term is just the variance of the HT estimator for the mean of imputed values via Algorithm \ref{algo-1}, and the $V_{2}$ term can be seen as compensation for the inherent randomness (induced into $S_{\text{NP}}$ by unknown sampling, regression estimation, etc.) resulting from PMM imputation. 

\begin{remark}(Regarding Theorem \ref{variance-pmm-proof})
\label{remark-4}
    When deriving the expression for $\mathbb{V}\left[\hat{\mu}\right]$, we conditioned variance on knowing the values of $y_{i}$ and $\hat{\nu}(i,t)$ for $i=1,\dotso,N$, $t=1,\dotso,k$. We can choose to do the same for nearest neighbour imputation in the spirit of \citet{yang2021integration} if we treat $k$ as a random variable and not a~hyperparameter we additionally have to condition on $k$. This gives us an exactly analogous expression for $V_{1}, V_{2}$ with matching variables $\hat{\nu}(i,t)$ constructed by the kNN algorithm from which exact variance for kNN imputation can be obtained (conditional on $k$ if it is a random variable). The additional variance component $V_{2}$ was not considered in \citet{yang2021integration}.
\end{remark}

As a corollary of Theorem \ref{variance-pmm-proof}, we have the following approximation for \eqref{mu-hat-definition-yhat-yhat-match} with unknown population size estimated by the usual $\hat{N}=\sum_{i\in S_{\text{P}}}\pi_{i}^{-1}$:
\begin{align}\label{variance-N-unknown}
    \mathbb{V}\left[\frac{N}{\displaystyle
    \sum_{r\in S_{\text{P}}}\pi_{r}^{-1}}\hat{\mu}\right]
    &\approx V_{1}+V_{2}+\mathbb{E}\left[\hat{\mu}\right]^{2}
    \frac{\mathbb{V}\left[\hat{N}\right]}{N^{2}}-
    2\mathbb{E}\left[\hat{\mu}\right]\frac{1}{N}\cov\left(\hat{N}, \hat{\mu}\right)\\
    \nonumber&=V_{1}+V_{2}+\frac{\mathbb{E}\left[\hat{\mu}\right]^{2}}{N^{2}}
    \sum_{i,j=1}^{N}\text{cov}\left(\1_{i\in S_{\text{P}}}\pi_{i}^{-1}, 
    \1_{j\in S_{\text{P}}}\pi_{j}^{-1}\right)\\
    \nonumber&-2\mathbb{E}\left[\hat{\mu}\right]\frac{1}{N^{2}}\sum_{i,j=1}^{N}\frac{1}{k}\sum_{t=1}^{k}
    \text{cov}\left(\1_{\{i\in S_{\text{P}}\}}\pi_{i}^{-1}, \pi_{j}^{-1}\1_{\{j\in S_{\text{P}}\}}
    Y_{\hat{\nu}(j, t)}\right)
\end{align}
Since $\displaystyle\hat{N}$ is unbiased for $N$, we obtain \eqref{variance-N-unknown} by Taylor series expansion of $f(x,y)=x/y$.

\subsection{Variance estimator}

\subsubsection{Partially analytic variance estimator}
Theorem \ref{variance-pmm-proof-est} provides a consistent estimator of the component $V_1$ of the variance of \eqref{mu-hat-definition-yhat-yhat-match}. The full proof is presented in Appendix \ref{proof-of-3}.

\begin{theorem}\label{variance-pmm-proof-est}
Let $z_{i}=\frac{1}{k}\sum_{t=1}^{k}Y_{\hat{\nu}(i,t)}$ denote the value imputed to the unit $i \in S_P$ and let $\mathcal{G}$ denote the $\sigma$-algebra as in the Proof of Theorem \ref{variance-pmm-proof}, i.e. $\sigma\left(\{Y_{i}, \bX_{i}, \hat{\nu}(i, t), \pi_{i}\}_{i\in U, t\in\{1,\dots,k\}}\right)$. Let
\begin{equation*}
    \mathbb{V}[\hat\mu \mid \mathcal{G}] = \frac{1}{N^2} \sum_{i=1}^N \sum_{j=1}^N \frac{\pi_{ij} - \pi_i\pi_j}{\pi_i\pi_j} z_i z_j
\end{equation*}
be the design variance of $\hat\mu$ conditional on $\mathcal{G}$. Assume that second order inclusion probabilities are known, $\displaystyle\left|(\pi_{ij}-\pi_{i}\pi_{j})\pi_{ij}^{-1}\right|$ is bounded for all $i,j$, assumptions of Theorems \ref{yhat-y-consistency-proof}, \ref{yhat-yhat-consistency-proof}, \ref{yhat-yhat-consistency-mis-specified} are satisfied for PMM~B, PMM~A under correct specification and PMM~A under misspecification, respectively, and the $\mathbb{E}[Y^4] < \infty$. If in addition the Horvitz-Thompson variance estimator is consistent (conditionally on $\mathcal{G}$) and $n_P \mathbb{V}[\hat\mu \mid \mathcal{G}] \overset{P}{\rightarrow} \gamma \in (0, \infty)$, then $\hat V_1$ defined as:
    \begin{equation}\label{pmm-known-N-var-est}
        \hat{V}_1 = 
        \frac{1}{N^{2}}\sum_{i \in S_{\text{P}}}\sum_{j \in S_{\text{P}}}
        \frac{\pi_{ij}-\pi_i\pi_j}{\pi_{ij}\pi_i\pi_j}
        \left(\frac{1}{k}\sum_{t=1}^{k}y_{\hat{\nu}(i, t)}\right)
        \left(\frac{1}{k}\sum_{t'=1}^{k}y_{\hat{\nu}(j, t')}\right),
    \end{equation}
    satisfies $\hat V_1 / V_1 \overset{P}{\rightarrow} 1$, and thus is a consistent estimator of $V_1$.
\end{theorem}

By Theorem \ref{variance-pmm-proof}, the total variance of the estimator can be decomposed into $V_1$ and $V_2$, where $V_2$ represents the additional randomness induced by the imputation procedure. In our PMM setting, the imputation error can be presented as:
\begin{equation*}
    \frac{1}{k}\sum_{t=1}^{k}Y_{\hat{\nu}(i, t)} - Y_i= \frac{1}{k}\sum_{t=1}^{k} \left(m(\bX_{\hat\nu (i,t)}, \bbeta_0) - m(\bX_i, \bbeta_0) \right) + \frac{1}{k}\sum_{t=1}^{k}(\varepsilon_{\hat\nu (i,t)} -\varepsilon_i)  + o_P(1),
\end{equation*}
where $o_P(1)$ accounts for the usage of $\hat\bbeta$ instead of $\bbeta_0$. In PMM A, matching is done using the predicted values, which implies that $m(\bX_{\hat\nu (i,t)}, \bbeta_0)$ is close to $m(\bX_i, \bbeta_0)$, and therefore the first term is small (asymptotically). On the other hand, PMM B matching is based on the observed outcomes, so the residual component $\varepsilon_{\hat\nu (i,t)} -\varepsilon_i$ may directly affect the selection of donors and does not necessarily vanish for fixed $k$. Consequently, $V_2$ may be non-negligible in finite samples. By Theorem \ref{variance-pmm-proof-est}, $\hat V_1$ is consistent only for the $V_1$. The total variance is targeted by $\hat V_1 + \hat V_2$ and $\hat V_1$ alone is consistent for the whole $\mathbb{V}[\hat\mu]$ only when $V_2/V_1 \rightarrow 0$, for example when $n_{P}/n_{NP} \rightarrow 0$.

The  $\hat{V}_1$ component is the Horvitz-Thompson variance estimator for the mean computed via imputed values. The term $\hat{V}_2$, which accounts for the additional component $V_2$ induced by the imputation, is given by:
\begin{equation*}
    \hat{V}_2 = \frac{1}{N^{2}}
    \sum_{i\in S_{\text{P}}}\sum_{j\in S_{\text{P}}}\pi_{ij}^{-1}
    \widehat{\text{cov}}\left(
    \frac{1}{k}\sum_{t=1}^{k}y_{\hat{\nu}(i, t)},
    \frac{1}{k}\sum_{t'=1}^{k}y_{\hat{\nu}(j, t')}\right).
\end{equation*}
It requires an estimator for covariance of imputed values, which we obtain via the mini-bootstrap described in Algorithm \ref{algo-2}. Fortunately, as we demonstrate in the simulation study, we can omit the $\hat{V}_{2}$ term for large $S_{\text{NP}}$ samples.  

If the population size $N$ is unknown and estimated from the probability sample, we can use the plug-in estimator for \eqref{variance-N-unknown} given by:
\begin{align}
    \widehat{\mathbb{V}}\left[\frac{N}{\displaystyle
    \sum_{r\in S_{\text{P}}}\pi_{r}^{-1}}\hat{\mu}\right]
    &=\frac{1}{\hat{N}^{2}}\sum_{i \in S_{\text{P}}}\sum_{j \in S_{\text{P}}}
    \frac{\pi_{ij}-\pi_i\pi_j}{\pi_{ij}\pi_i\pi_j}
    \left(\frac{1}{k}\sum_{t=1}^{k}y_{\hat{\nu}(i, t)}\right)
    \left(\frac{1}{k}\sum_{t'=1}^{k}y_{\hat{\nu}(j, t')}\right)\\
    \nonumber&+\frac{1}{\hat{N}^{2}}
    \sum_{i\in S_{\text{P}}}\sum_{j\in S_{\text{P}}}\pi_{ij}^{-1}
    \widehat{\text{cov}}\left(
    \frac{1}{k}\sum_{t=1}^{k}y_{\hat{\nu}(i, t)},
    \frac{1}{k}\sum_{t'=1}^{k}y_{\hat{\nu}(j, t')}\right)\\
    \nonumber&+\frac{\hat{\mu}^{2}}{\hat{N}^{2}}
    \sum_{i,j\in S_{\text{P}}}\frac{\pi_{ij}-\pi_{i}\pi_{j}}{\pi_{ij}\pi_{i}\pi_{j}}
    -2\frac{\hat{\mu}}{\hat{N}^{2}}
    \sum_{i,j\in S_{\text{P}}}\frac{\pi_{ij}-\pi_{i}\pi_{j}}{\pi_{ij}\pi_{i}\pi_{j}} \left(\frac{1}{k}\sum_{t=1}^{k}y_{\hat{\nu}(i, t)}\right) ,
\end{align}
\noindent where the first two lines refer to estimators for $V_{1}$ and $V_{2}$ under estimated population size, and the last line is a~correction for randomness of $\hat{N}$. 

Unfortunately, the $V_{2}$ term cannot be estimated directly from samples $S_{\text{NP}}$ and $S_{\text{P}}$; therefore, we propose a simple "mini-bootstrap" to estimate $\text{cov}\left(\frac{1}{k}\sum_{t=1}^{k}y_{\hat{\nu}(i, t)},\frac{1}{k}\sum_{t'=1}^{k}y_{\hat{\nu}(j, t')}\right)$. The procedure is explained in Algorithm \ref{algo-2}.

\begin{algorithm}[ht]
\small
\caption{Non-parametric mini-bootstrap estimator for covariance terms}
\label{algo-2}\DontPrintSemicolon
\nlset{1} Sample $n_{\text{NP}}$ units from $S_{\text{NP}}$ with replacement to create $S_{\text{NP}}'$ (if pseudo-weights are present inclusion probabilities should be proportional to their inverses).\;
\nlset{2} Estimate regression model $\mathbb{E}[Y|\bX]=m(\bX, \bbeta)$ based on $j\in S_{\text{NP}}'$ from step \texttt{1}.\;
\nlset{3} Compute $\hat{\nu}'(i,t)$ for $t=1,\dots,k, i\in S_{\text{P}}$ using estimated $m(\bx', \cdot)$ and $\{(y_{j},\bx_{j})| j\in S_{\text{NP}}'\}$.\;
\nlset{4} Compute $\frac{1}{k}\sum_{t=1}^{k}y_{\hat{\nu}'(i,t)}$ using $Y$ values from $S_{\text{NP}}'$.\;
\nlset{5} Repeat steps \texttt{1}-\texttt{4} $M$ times (we set $M=50$ in our simulations).\;
\nlset{6} Estimate covariance between imputed values for each pair $i,j\in S_{\text{P}}$ using constructed pseudo-sample with values from \texttt{4}.\;
\end{algorithm}

The $V_1$ component follows from the design-based variance decomposition, conditional on the realised imputed values. The $V_2$ component is different: it reflects the additional randomness induced by the source sample and by the fixed-$k$ matching map. We therefore interpret Algorithm~\ref{algo-2} as a computational finite-sample correction for the covariance terms in $V_2$, rather than as a generally consistent bootstrap estimator. We do not claim general consistency of this correction. Its use is heuristically motivated when the source sample behaves approximately as independent weighted draws, donor reuse is diffuse, and the matching functional is stable. In the simulation study, we report results both with and without this correction.

The intuitive reason why Algorithm \ref{algo-2} may be an appropriate bootstrap for $\hat{V}_{2}$ is that, as per equation \eqref{V-2-expression}, $V_{2}$ is just the variance of the mean of imputed values for the whole population. Since imputation for the whole population can be done with only the $S_{\text{NP}}$ sample (if we know $\bX$ values for the entire population), the only source of randomness that is relevant to $V_{2}$ is the random selection of units into $S_{\text{NP}}$. Therefore, we only need to account for sampling in $S_{\text{NP}}$ and use the probability sample $S_{\text{P}}$ to estimate the $\frac{1}{N}\sum_{i=1}^{N}\frac{1}{k}\sum_{t=1}^{k}y_{\hat{\nu}(i, t)}$ using the usual weighted estimator and compute the variance of estimates obtained via Algorithm \ref{algo-2}. This also explains why results from Algorithm \ref{algo-2} are somewhat conservative since we are essentially estimating
\begin{equation*}
    \mathbb{V}\left[\frac{1}{N}\sum_{i=1}^{N}
    \frac{1}{k}\sum_{t=1}^{k}y_{\hat{\nu}(i, t)}\right],
\end{equation*}
by an estimator for:
\begin{equation*}
    \mathbb{V}\left[\frac{1}{N}\sum_{i\in S_{\text{P}}}
    \frac{1}{\pi_{i}}\frac{1}{k}\sum_{t=1}^{k}y_{\hat{\nu}(i, t)}\right],
\end{equation*}
which induces more randomness.

\subsection{Bootstrap variance estimator}\label{sec-bootstrap-mini-included}

In this section, we provide a bootstrap variance estimator for the mean. Since we estimate $V_2$ using the ``mini-bootstrap'' described by Algorithm \ref{algo-2}, we need to estimate $V_1$ using the appropriate bootstrap approach for probability samples.

\begin{algorithm}[ht!]
\small
\caption{Bootstrap variance estimator}
\label{algo-3}\DontPrintSemicolon
\nlset{1} Sample with replacement $n_{\text{NP}}$ units from $S_{\text{NP}}$ to create $S_{\text{NP}}'$ (if pseudo-weights are present, inclusion probabilities should be proportional to their inverses).\;
\nlset{2} Sample with replacement $n_{\text{P}}$ units from $S_{\text{P}}$ according to the sampling design to create $S_{\text{P}}'$.\;
\nlset{3} Estimate regression model parameters $\mathbb{E}[Y|\bX=\bx']=m(\bx', \bbeta)$ based on $j\in S_{\text{NP}}'$ from step \texttt{1}.\;
\nlset{4} Follow steps 2A/2B-3 from Algorithm \ref{algo-1} depending on the imputation method ($\hat{y}-\hat{y}$ or $\hat{y}-y$) to compute $\hat{\mu}$ in current iteration.\;
\nlset{5} Repeat steps \texttt{1}-\texttt{4} $L$ times (we set $L=500$ in our simulations).\;
\nlset{6} Estimate $V_{2}$ term using Algorithm \ref{algo-2}.\;
\nlset{7} Estimate variance $V_{1}$ term as $\hat{V}_{1}=\frac{1}{L - 1} \sum_{i=1}^{L} \left(\hat{\mu}_{i} - \bar{\mu}\right)^{2}$ where $\bar{\mu} = \frac{1}{L}\sum_{i=1}^{L} \hat{\mu}_{i}$ and final bootstrap variance as $\hat{V} = \hat{V}_{1} + \hat{V}_{2}$.\;
\end{algorithm}

In the simulation study, we compare the analytic bootstrap approaches and show that both estimators have the assumed nominal coverage rate.

\section{Simulation study}\label{sec-sim}

We generate a finite population of size $N=10^{5},$ three random variables $X_1, X_2$ and $X_3$ independently from the $\mathcal{N}(2,1)$ and $\varepsilon\sim \mathcal{N}(0,1)$ and assume the following models: 1) $y_{1i}=1+2x_{1i}+\varepsilon_i$; 2) $y_{2i}=-1 + x_{1i} + x_{2i} + x_{3i} + \varepsilon_i$; and 3) $y_{3i}=-10 + x_{1i}^2 + x_{2i}^2 + x_{3i}^2 +\varepsilon_i$. We consider two sample sizes for the non-probability sample $n_{\text{NP}}=500$ and $n_{\text{NP}}=1,000$. Sample $S_{\text{NP}}$ is drawn from the population using simple random sampling without replacement (SRSWOR) within two strata created by $x_{1i} \leq 2$ and $x_{1i} >2$ with the following sample sizes: $n_{strata1}=0.7 n_{\text{NP}}$ and $n_{strata2}=0.3n_{\text{NP}}$, respectively. For the probability sample, we assume a fixed sample size $n_{\text{P}}=500$ and SRSWOR from the population.  

The goal of the study is to verify the performance of the proposed estimators in comparison to two existing alternatives for mass imputation: nearest neighbour (denoted as NN) and regression prediction (denoted as GLM). We verify the effect of the number of covariates as well as model misspecification. 

For all target variables, we assume a correctly specified model as defined above. In addition, for $Y_3$ we use a misspecified model (i.e. linear regression with $X_1$ and $X_2$ only). In each of $500$ simulation runs and for every $y$, we calculated 

\begin{itemize}
    \item The naive estimator (sample mean) from sample $A$, given by $\hat{\mu}=n_{\text{NP}}^{-1}\sum_{i \in S_{\text{NP}}}y_{i}$.
    \item The mass imputation estimator using linear regression denoted as \texttt{GLM}. 
    \item The NN mass imputation estimators. We studied the impact of $k$ using $k=1$ and $k=5$ which is denoted as \texttt{NN1} and \texttt{NN5}.
    \item Two PMM mass imputation estimators given by \eqref{mu-hat-definition-yhat-yhat-match} using $k=1$ denoted as \texttt{PMM1A} for $\hat{y}-\hat{y}$ and \texttt{PMM1B} for $\hat{y}-y$ matching. 
    \item Two PMM mass imputation estimators given by \eqref{mu-hat-definition-yhat-yhat-match} using $k=5$ denoted as \texttt{PMM5A} for $\hat{y}-\hat{y}$ and \texttt{PMM5B} for $\hat{y}-y$ matching. 
\end{itemize}

Three results are reported in this section. Firstly, we report Monte Carlo bias (Bias), standard error (SE), root mean square error (RMSE), and empirical coverage rate (CR) of 95\% confidence intervals using the analytical variance estimator based on $R=500$ simulations for each $Y$ variable: $\text{Bias}  = \bar{\hat{\mu}} - \mu$, $\text{SE}= \sqrt{\sum_{r=1}^R \left(\hat{\mu}^{(r)} - \bar{\hat{\mu}}\right)^2/(R-1)}$, and $\text{RMSE} = \sqrt{ \text{Bias}^2 + \text{SE}^2}$, where $\bar{\hat{\mu}} = \sum_{r=1}^{R}\hat{\mu}^{(r)} / R$ and $\hat{\mu}^{\left(r\right)}$ is an estimate of the mean in the $r$-th replication. Secondly, we compare how the CR changes if we drop $\hat{V}_2$ from the NN and PMM variance estimators. Finally, we compare analytic and bootstrap estimators for the PMM estimators discussed in this section.

Table \ref{tab-simulation-results} presents the main results of the simulation study. For the first variable ($Y_1$), where the model is correctly specified, values of both the bias and the RMSE are similar to the GLM estimator proposed by \citet{kim_combining_2021}. As expected, the results for the NN and PMM estimators are exactly the same. The main difference between the proposed PMM estimators is the standard error, which is larger for $\hat{y}-\hat{y}$ imputation compared to the standard $\hat{y}-y$ imputation. This relationship is particularly visible for the larger sample ($n=1000$). In the case of $Y_1$, there is no significant difference between the proposed PMM $\hat{y}-y$ and the GLM estimator. 

\begin{center}
    Table \ref{tab-simulation-results} around here.
\end{center}

As regards the second variable $Y_2$, a similar pattern can be observed as for $Y_1$ with one exception. The larger the number of nearest neighbours, the higher the bias observed for the NN estimator as we use 3 variables  (as proved by \citet{yang2021integration} asymptotic bias of the NN estimator increases with the number of variables). For the PMM B estimator increasing the number of nearest neighbours has no significant effect on the bias or the standard error. The empirical coverage rates for all estimators are close to the nominal 95\%. 

The main differences can be observed for the last variable $Y_3$ (non-linear). If we use correctly specified transformations and linear regression for all estimators the bias is negligible for the GLM and PMM estimators but not for the NN estimator.  The PMM A with $\hat{y}-\hat{y}$ fitting and the GLM are now characterised by the lowest bias and a CR close to the nominal 95\%. This is in line with robustness property of $\hat{y}-\hat{y}$ matching established in Section \ref{sec-main-misspecified}.

As expected, when the model is misspecified (rows under \textit{misspecified}), i.e. we use linear regression with $x_1$ and $x_2$ only, all estimators are characterised with bias. However, the bias is the lowest for the PMM A estimator. Interestingly, increasing $k$ for the NN leads to an increase of bias. On the other hand, the PMM B estimators are characterised with the lowest variance and thus lowest RMSE (along with PMM5A). All estimators have CR close to the nominal 95\%.

\begin{center}
    Table \ref{tab-res2} around here.
\end{center}

Table \ref{tab-res2} shows two results. The first part of the table shows the effect of the second component $V_2$ on the variance estimator of the mean. The second compares the analytic and bootstrap variance estimators as discussed in the previous section. As suggested by theory, the contribution of $V_2$ component decreases as $n_{NP}$ increases for both NN and PMM estimators. The NN estimator for the $Y_3$ variable is characterised with the lowest CR in both scenarios. The second part suggests that the proposed analytical and bootstrap variance estimation approaches yield the same results for all study variables and sample sizes. As expected, the number of nearest neighbours does not affect the CR as only small differences are observed. 

In the Appendix we provide additional results: 1) choosing $k$ number of nearest neighbours does not improve coverage but lowers MSE (as was also shown by \citet{yang2021integration} for the NN estimator, see section \ref{sec-appen-sim-1}); 2) variable selection for the GLM and PMM estimators using the SCAD penalty (as in \citet{yang_doubly_2020}) improves estimates and the coverage rate and should be considered in applications (see section \ref{sec-appen-sim-2}); 3) non-parametric regression using a local linear model can be employed, although in practice better non-parametric methods ought to be considered (see section \ref{sec-appen-sim-3}); 4) the proposed estimators perform similarly to the GLM estimator when the positivity assumption is violated with some exceptions when a non-linear model is considered (see section \ref{sec-appen-sim-4}); and finally 5) the initial result on the multiply robust procedure proposed by \citet{chen_note_2021} seems to work well: for the PMM~A estimator it provides results comparable to standard matching (i.e. without the property of being multiply robust), while for the PMM~B estimator it can substantially reduce the bias for the non-linear variable (see section \ref{sec-appen-sim-5}). This, however, requires further studies as no theory has been developed for the multiply robust PMM estimator. 

\section{Empirical study}\label{sec-empirical}

In this section we present an attempt to integrate administrative and survey data about job vacancies for the end of 2022Q2 in Poland. The aim was to estimate the share of vacancies aimed at Ukrainian workers. We defined our target variable as follows: $Y$ indicates \textit{whether the vacancy has been translated into Ukrainian}. As the survey only provides information about the number of vacancies and not specific vacancies, our models were based on the independent variable $y$ defined as the share of vacancies translated into Ukrainian calculated separately for each unit. We present a~brief description of the datasets, but we encourage readers to read the full description of the dataset and the relationships between the target and auxiliary variables as described in \citet{beresewicz2025quantile}.

The first source we used is the Job Vacancy Survey (JVS, also known as the Labour Demand Survey), which is a~stratified sample of 100,000 units, with a~response rate of about 60\% ($S_{\text{P}}$). The survey population consists of companies and their local units with 1 or more employees. The sampling frame includes information about NACE (19 levels), region (16 levels), sector (2 levels), size (3 levels) and the number of employees according to administrative data integrated by Statistics Poland (RE). The JVS sample contains 304 strata created separately for enterprises with up to 9 employees and those with more than 10 employees \citep[cf.][]{jvs-meth}. Of the 60,000 responding units, around 7,000 reported at least one vacancy. Our target population included units with at least one vacancy; according to the survey, there were between 38,000 and 43,000 of such units at the end of 2022Q2.

The second source is the Central Job Offers Database (CBOP), which is a~register of all vacancies submitted to Public Employment Offices (PEOs, our $S_{\text{NP}}$). CBOP is available online and can be accessed via API. CBOP includes all types of contracts and jobs outside Poland, so data cleaning was carried out to align the definition of a vacancy with that used in the JVS. CBOP data collected via API include information about whether a~vacancy is outdated (e.g. 17\% of vacancies were outdated when downloaded at the end of  2022Q2). CBOP also contains information about unit identifiers (REGON and NIP), so we were able to link units to the sampling frame to obtain auxiliary variables with the same definitions as those used in the survey (24\% of records contained no identifier because the employer has the right to withhold this information). The final CBOP dataset contained about 8,500 units included in the sampling frame.

The following variables were considered: Region (16 levels), NACE (19 levels), Sector (2 levels), Size (3 levels), $\log$(RE), $\log$(vac) (the number of vacancies), $I$(\#~vacancies = 1) (whether employer seeks only  one person). We considered 5 estimators in the empirical study (for all models linear regression is assumed and neighbors set to $k=5$):

\begin{itemize}
    \item Mass imputation (MI) estimators: GLM \citep{kim_combining_2021}, NN1 (with all variables), NN2 ($\log$(RE) and $\log$(vac) only), PMM A, PMM B,
    \item Inverse probability weighting (IPW) estimator without (IPW MLE) and with calibration constraints \citep[cf.][]{chen2020doubly} (IPW GEE),
    \item Doubly robust (DR) estimator with IPW MLE.
\end{itemize}

Variance was estimated using the following bootstrap approach: 1) JVS sample was resampled using a~stratified bootstrap approach, 2) CBOP was resampled using simple random sampling with replacement. This procedure was repeated $B=500$ times with the \textit{mini-bootstrap} for $V_2$ as described in Section \ref{sec-bootstrap-mini-included}.  Table \ref{tab-results-emp} shows point estimates (denoted as Point), bootstrap standard errors (denoted as SE), the coefficients of variation (CV) and 95\% confidence intervals. For NN and PMM estimators we report results with and without the mini-bootstrap for $V_2$.

\begin{center}
    Table \ref{tab-results-emp} around here.
\end{center}

As can be seen in Table \ref{tab-results-emp}, the point estimates produced by all the estimators in the study are at a similar level and fluctuate around 22\%, with the exception of the Na\"ive estimator (20.51\%), for which the estimated share of job vacancies aimed at Ukrainians is lower, and the NN1 estimator (24.01\%), for which it is somewhat higher. CV values for the discussed estimators are similar, with the lowest values for the mass imputation estimators. For the NN and PMM estimators the mini-bootstrap for $V_2$ noticeably increases the standard errors and CV, while leaving the point estimates unchanged; accounting for $V_2$ increases the variance of the matching estimators by roughly 40--60\% (e.g. the PMM CV rises from 5.69 to 6.77), so the matching uncertainty is non-negligible in practice. Among the mass imputation estimators, GLM is the most efficient, with smaller standard errors than IPW, while the NN and PMM estimators are less efficient than IPW once $V_2$ is taken into account; the IPW (GEE), DR and NN2 estimates agree closely at around 21--22\%. Results for PMM A and PMM B were indistinguishable, so a single PMM estimator is reported.

\section{Conclusions}

We have discussed the asymptotic properties of two PMM estimators that can be used to integrate data from non-probability and probability samples. We have proved their consistency and derived variance decomposition, with an analytical estimator for the design-based component and a bootstrap correction for the remainder. The results obtained in the study regarding the PMM~B hold not only for parametric but also for non-parametric models used for matching. The variance estimators can be applied to the NN estimator for non-probability samples.

A possible application of $\hat{y}-y$ matching outside of just being more efficient under heavy assumptions is that we can consider mixed matching with respect to the distance function:
\begin{equation*}
    d(i, j)=\frac{1}{a_{N}}\left|m\left(\bx_{i}, \hat{\bbeta}\right)-y_{j}\right|+
    \left|m\left(\bx_{i}, \hat{\bbeta}\right)-m\left(\bx_{j}, \hat{\bbeta}\right)\right|,
\end{equation*}
or:
\begin{equation*}
    d(i, j)=\left|\frac{1}{a_{N}}\left(m\left(\bx_{i}, \hat{\bbeta}\right)-y_{j}\right)+
    m\left(\bx_{i}, \hat{\bbeta}\right)-m\left(\bx_{j}, \hat{\bbeta}\right)\right|,
\end{equation*}

\noindent where $a_{N}\rightarrow\infty$ as $N\rightarrow\infty$ but slowly so that the new estimator retains asymptotic properties of $\hat{y}-\hat{y}$ matching but is closer to $\hat{y}-y$ matching in finite samples. Some initial simulations gave very optimistic results. The performance of such estimators and the optimal choice of a~sequence $a_{n}$ will be subject to further research.

In the simulation study, we have shown that the PMM estimator based on $\hat{y}-\hat{y}$ matching is robust to model misspecification and has a lower bias than existing alternatives. This is in line with results on PMM estimators for survey non-response or statistical matching. However, this gain is achieved at the expense of an increase in variance, but only for linear models. In additional simulation studies presented in the appendix, we verified the use of PMM estimators for different settings. 

Simulation in the Appendix \ref{sec-appen-sim-1} shows initial results on a~dynamic selection of $k$ hyperparameter. The results are promising, but one should be aware that minimizing the estimated variance way of choosing $k$ though the variance should then be adjusted due to randomness in $k$ via:
\begin{equation*}
    \mathbb{V}\left[\hat{\mu}\right]=
    \mathbb{E}\left[\mathbb{V}\left[\left.\hat{\mu}\right|k\right]\right]+
    \mathbb{V}\left[\mathbb{E}\left[\left.\hat{\mu}\right|k\right]\right]
    =\sum_{t=1}^{\infty}(V_{1}(t)+V_{2}(t))\mathbb{P}\left[k=t\right]+
    \mathbb{V}\left[\mathbb{E}\left[\left.\hat{\mu}-\mu\right|k\right]\right],
\end{equation*}

\noindent which could be very difficult and could be the subject of further research. Sensitivity analysis via CV is always a viable alternative.

A natural extension of the study is that we can maintain consistency if the design weights $d_i$ are replaced with calibration weights, which is often the case when working with survey sample datasets. Another problem worth investigating is whether we can replace probability convergence in $\lVert\cdot\rVert_{\infty}$ with weaker convergence for estimators of $\hat{m}$ from  Remark \ref{remark-2}. In finite dimensional spaces all norms and topologies induced by them are equivalent, so it only matters for non-parametric regression models. Further studies may focus on the best method for estimating $V_{2}$, for which we have proposed a "mini" bootstrap.

\printbibliography

\clearpage

\begin{table}[ht!]
    \centering
    \caption{Two sample setting.}
    \label{tab-two-sources}
    \begin{tabular}{llccc}
    \hline
     Sample    & ID & \begin{tabular}[c]{@{}l@{}}Sample weight\\ $d=\pi^{-1}$\end{tabular}& Covariates $\bx$ & Study variable $y$ \\
    \hline
     Non-probability sample ($S_{\text{NP}})$   & 1 & ? & $\checkmark$ & $\checkmark$ \\
     & $\vdots$ & ? & $\vdots$ & $\vdots$ \\
     & $n_{\text{NP}}$ & ? & $\checkmark$ & $\checkmark$ \\
     Probability sample ($S_{\text{P}}$)      & 1 & $\checkmark$ & $\checkmark$ & ?\\
     & $\vdots$ & $\vdots$ & $\vdots$ & ?\\ 
     & $n_{\text{P}}$ & $\checkmark$ & $\checkmark$ & ?\\                                     
    \hline     
    \end{tabular}
\end{table}

\clearpage

\begin{table}[ht]
\centering
\scriptsize
\caption{Simulated Bias, SE, RMSE (multiplied by 100) and CR of 5 estimators}
\label{tab-simulation-results}
\begin{tabular}{lrrrrrrrrrrrr}
  \hline
  Estimator & \multicolumn{4}{c}{$Y_1$} & \multicolumn{4}{c}{$Y_2$} & \multicolumn{4}{c}{$Y_3$} \\ 
   & Bias & SE & RMSE & CR & Bias & SE & RMSE & CR & Bias & SE & RMSE & CR \\ 
  \hline
  \multicolumn{13}{c}{$n_{\text{NP}}=500$}  \\
  \hline
  naive & -64.61 & 5.30 & 64.82 &  & -32.32 & 6.11 & 32.89 &  & -128.93 & 21.60 & 130.72 &  \\ 
   \hline
  \multicolumn{13}{c}{\textit{correctly specified}}  \\
  \hline
  GLM & 0.10 & 9.64 & 9.64 & 95.60 & -0.17 & 8.60 & 8.60 & 94.00 & 0.23 & 33.21 & 33.21 & 93.60 \\ 
  NN1 & 0.35 & 10.61 & 10.62 & 94.40 & -1.84 & 9.36 & 9.54 & 95.00 & -19.84 & 32.10 & 37.74 & 92.20 \\ 
  NN5 & 0.05 & 9.76 & 9.76 & 95.40 & -2.64 & 8.29 & 8.71 & 94.00 & -37.66 & 30.64 & 48.55 & 77.20 \\ 
  PMM1A & 0.35 & 10.61 & 10.62 & 94.20 & -0.04 & 9.91 & 9.91 & 95.80 & -0.28 & 33.79 & 33.79 & 93.60 \\ 
  PMM1B & 0.06 & 9.63 & 9.63 & 95.60 & -0.19 & 8.59 & 8.60 & 93.40 & -0.48 & 33.07 & 33.07 & 93.80 \\ 
  PMM5A & 0.05 & 9.76 & 9.76 & 95.40 & -0.24 & 8.86 & 8.86 & 94.40 & -1.85 & 32.89 & 32.94 & 93.20 \\ 
  PMM5B & -0.03 & 9.60 & 9.60 & 95.20 & -0.24 & 8.58 & 8.59 & 93.60 & -1.82 & 32.80 & 32.85 & 93.60 \\ 
 \hline
  \multicolumn{13}{c}{\textit{misspecified}}  \\
  \hline
  GLM   &  &  &  &  &  &  &  &  & -10.20 & 30.34 & 32.01 & 92.80 \\ 
  NN1   &  &  &  &  &  &  &  &  & -6.14 & 40.04 & 40.51 & 94.20 \\ 
  NN5   &  &  &  &  &  &  &  &  & -13.57 & 33.01 & 35.69 & 90.80 \\ 
  PMM1A &  &  &  &  &  &  &  &  & -3.53 & 40.30 & 40.46 & 95.60 \\ 
  PMM1B &  &  &  &  &  &  &  &  & -9.65 & 30.27 & 31.77 & 92.60 \\ 
  PMM5A &  &  &  &  &  &  &  &  & -4.05 & 33.67 & 33.91 & 95.40 \\ 
  PMM5B &  &  &  &  &  &  &  &  & -9.38 & 30.23 & 31.66 & 92.80 \\ 
  
  \hline
  \multicolumn{13}{c}{$n_{\text{NP}}=1,000$}  \\
  \hline
  naive & -63.50 & 7.03 & 63.89 &  & -31.49 & 8.30 & 32.57 &  & -127.37 & 28.89 & 130.61 &  \\ 
\hline
  \multicolumn{13}{c}{\textit{correctly specified}}  \\
  \hline
  GLM   & 0.43 & 10.20 & 10.21 & 93.80 & 0.10 & 9.22 & 9.22 & 94.80 & 0.55 & 33.40 & 33.40 & 93.40 \\ 
  NN1   & 0.01 & 11.37 & 11.37 & 95.00 & -1.86 & 10.14 & 10.31 & 95.00 & -28.20 & 32.22 & 42.82 & 86.80 \\ 
  NN5   & -0.09 & 10.41 & 10.41 & 95.40 & -3.41 & 8.85 & 9.49 & 93.80 & -53.77 & 29.78 & 61.47 & 57.40 \\ 
  PMM1A & 0.01 & 11.37 & 11.37 & 95.20 & 0.08 & 10.62 & 10.62 & 95.80 & -1.11 & 33.42 & 33.43 & 94.20 \\ 
  PMM1B & 0.35 & 10.19 & 10.19 & 93.60 & 0.06 & 9.20 & 9.20 & 95.00 & -0.82 & 33.15 & 33.16 & 93.40 \\ 
  PMM5A & -0.09 & 10.41 & 10.41 & 95.60 & -0.28 & 9.41 & 9.42 & 95.00 & -3.65 & 32.82 & 33.02 & 93.60 \\ 
  PMM5B & 0.11 & 10.14 & 10.14 & 93.60 & -0.05 & 9.18 & 9.18 & 94.80 & -3.55 & 32.77 & 32.96 & 93.40 \\
  \hline
  \multicolumn{13}{c}{\textit{misspecified}}  \\
  \hline
  GLM    &  &  &  &  &  &  &  &  & -12.59 & 33.15 & 35.46 & 92.00 \\ 
  NN1    &  &  &  &  &  &  &  &  & -10.99 & 40.59 & 42.05 & 93.80 \\ 
  NN5    &  &  &  &  &  &  &  &  & -24.29 & 34.42 & 42.13 & 87.20 \\ 
  PMM1A  &  &  &  &  &  &  &  &  & -5.54 & 42.77 & 43.12 & 96.40 \\ 
  PMM1B  &  &  &  &  &  &  &  &  & -11.90 & 33.08 & 35.15 & 92.60 \\ 
  PMM5A  &  &  &  &  &  &  &  &  & -8.24 & 36.85 & 37.76 & 95.60 \\ 
  PMM5B  &  &  &  &  &  &  &  &  & -11.56 & 33.02 & 34.99 & 92.20 \\ 
  \hline
\end{tabular}
\end{table}

\clearpage

\begin{table}[ht!]
\centering
\caption{Empirical coverage rate intervals for two cases: the effect of $V_2$ on the variance estimator and analytic and bootstrap variance estimators.}
\label{tab-res2}
\begin{tabular}{lrrrrrr}
  \hline
  & \multicolumn{2}{c}{$Y_1$} &  \multicolumn{2}{c}{$Y_2$} & \multicolumn{2}{c}{$Y_3$} \\
  \hline
  \hline
  \multicolumn{7}{c}{Effect of $V_2$ on the variance estimator and its empirical coverage rate}  \\
  \hline
  \hline
  & $V_1$ & $V_1+V_2$ & $V_1$ & $V_1+V_2$ & $V_1$ & $V_1+V_2$ \\ 
  \hline
  \multicolumn{7}{c}{$n_{\text{NP}}=500$}  \\
  \hline
  NN1   & 91.00 & 95.00 & 91.60 & 95.00 & 84.40 & 86.80 \\ 
  NN5   & 92.20 & 95.40 & 84.00 & 93.80 & 53.60 & 57.40 \\ 
  PMM1A & 91.00 & 95.20 & 89.60 & 95.80 & 94.80 & 94.20 \\ 
  PMM1B & 94.00 & 93.60 & 92.00 & 95.00 & 95.80 & 93.40 \\ 
  PMM5A & 92.20 & 95.60 & 92.00 & 95.00 & 95.40 & 93.60 \\ 
  PMM5B & 93.60 & 93.60 & 92.40 & 94.80 & 95.80 & 93.40 \\ 
  \hline
  \multicolumn{7}{c}{$n_{\text{NP}}=1,000$}  \\
  \hline
  NN1   & 93.60 & 94.40 & 92.00 & 95.00 & 91.80 & 92.20 \\ 
  NN5   & 94.40 & 95.40 & 92.00 & 94.00 & 76.40 & 77.20 \\ 
  PMM1A & 93.60 & 94.20 & 92.80 & 95.80 & 95.60 & 93.60 \\ 
  PMM1B & 95.80 & 95.60 & 93.60 & 93.40 & 94.80 & 93.80 \\ 
  PMM5A & 94.40 & 95.40 & 93.60 & 94.40 & 95.80 & 93.20 \\ 
  PMM5B & 95.60 & 95.20 & 93.40 & 93.60 & 95.60 & 93.60 \\ 
  \hline
  \hline
  \multicolumn{7}{c}{Comparison of analytic and bootstrap variance estimators}  \\
  \hline
  \hline
   & Analytic & Bootstrap & Analytic & Bootstrap & Analytic & Bootstrap \\ 
   \hline
  \multicolumn{7}{c}{$n_{\text{NP}}=500$}  \\
  \hline
  NN1   & 95.00 & 97.20 & 95.00 & 97.00 & 86.80 & 86.20 \\ 
  NN5   & 95.40 & 97.40 & 93.80 & 96.80 & 57.40 & 59.60 \\ 
  PMM1A & 95.20 & 96.60 & 95.80 & 97.80 & 94.20 & 93.80 \\ 
  PMM1B & 93.60 & 96.00 & 95.00 & 97.00 & 93.40 & 93.20 \\ 
  PMM5A & 95.60 & 97.00 & 95.00 & 98.00 & 93.60 & 93.60 \\ 
  PMM5B & 93.60 & 95.80 & 94.80 & 97.20 & 93.40 & 94.40 \\ 
  \hline
  \multicolumn{7}{c}{$n_{\text{NP}}=1,000$}  \\
  \hline
  NN1   & 94.40 & 96.40 & 95.00 & 96.80 & 92.20 & 91.00 \\ 
  NN5   & 95.40 & 96.80 & 94.00 & 96.60 & 77.20 & 77.60 \\ 
  PMM1A & 94.20 & 96.60 & 95.80 & 96.80 & 93.60 & 94.00 \\ 
  PMM1B & 95.60 & 96.60 & 93.40 & 95.80 & 93.80 & 93.00 \\ 
  PMM5A & 95.40 & 96.40 & 94.40 & 97.00 & 93.20 & 93.60 \\ 
  PMM5B & 95.20 & 96.20 & 93.60 & 95.40 & 93.60 & 93.60 \\ 
   \hline
\end{tabular}
\end{table}

\clearpage

\begin{table}[ht!]
\centering
\caption{Estimates of the share of job vacancies aimed at Ukrainians at the end of 2022Q2 in Poland} 
\label{tab-results-emp}
\begin{tabular}{lrrrrr}
  \hline
Estimator & Point & SE & CV & 2.5\% & 97.5\% \\ 
  \hline
Na\"ive & 20.51 & -- & -- & -- & -- \\ 
   \hline
IPW (MLE) & 22.88 & 0.81 & 3.52 & 21.30 & 24.46 \\ 
  IPW (GEE) & 21.75 & 0.87 & 4.01 & 20.04 & 23.46 \\ 
  GLM & 22.68 & 0.60 & 2.64 & 21.51 & 23.85 \\ 
  DR & 21.28 & 0.70 & 3.31 & 19.90 & 22.67 \\ 
   \hline
\multicolumn{6}{c}{with 'mini-bootstrap' for $V_2$} \\
\hline
NN1 & 24.01 & 1.52 & 6.34 & 21.03 & 27.00 \\ 
  NN2 & 21.66 & 1.33 & 6.14 & 19.06 & 24.27 \\ 
  PMM & 23.42 & 1.59 & 6.77 & 20.31 & 26.53 \\ 
   \hline
\multicolumn{6}{c}{without 'mini-bootstrap' for $V_2$} \\
\hline
NN1 & 24.01 & 1.24 & 5.16 & 21.58 & 26.44 \\ 
  NN2 & 21.66 & 1.05 & 4.85 & 19.60 & 23.72 \\ 
  PMM & 23.42 & 1.33 & 5.69 & 20.81 & 26.03 \\ 
   \hline
\end{tabular}
\end{table}

\appendix
\clearpage

\makeatletter
\renewcommand{\partname}{}
\renewcommand{\thepart}{}
\makeatother

\part{Appendix}
\pagenumbering{arabic}
\localtableofcontents

\clearpage

\section{Assumptions}\label{app-assumptions}
\subsection{Regarding consistency of regression}\label{app-consistency-reg}

We do not strictly require that $\mathbb{P}[Y\in dy|\bX,\delta]=\mathbb{P}[Y\in dy|\bX]$ almost surely but in practice checking the consistency of regression may require checking this condition, as one reliable way of assessing consistency of $\hat{\bbeta}$ (see e.g. the discussion in \citet{kim_combining_2021}) is checking whether:
\begin{itemize}
    \item[(A1.1)] $\mathbb{P}[Y\in dy|\bX,\delta]=\mathbb{P}[Y\in dy|\bX]$ holds and
    \item[(A1.2)] $\hat{\bbeta}$ is a unique solution to score equations:
    \begin{equation*}
        \frac{1}{n_{\text{NP}}}\sum_{i\in S_{\text{NP}}}\left(y_{i}-m\left(\bx_{i}, \hat{\bbeta}\right)\right)\boldsymbol{h}\left(\bx_{i}, \hat{\bbeta}\right)=\boldsymbol{0},
    \end{equation*}
    for some $\boldsymbol{h}:\mathbb{R}^{p}\times\mathbb{R}^{p}\rightarrow\mathbb{R}^{p}$ chosen in such a way that some more technical assumptions (cf. \cite{Tsiatis2006}) are met, namely for $\bbeta$ is some neighbourhood of $\bbeta_{0}$:
    \begin{enumerate}
        \item $\displaystyle\mathbb{E}_{\bbeta}\left[\left(Y-m(\bX,\bbeta)\right)\boldsymbol{h}\left(\bX, \bbeta\right)\right]=\boldsymbol{0}$ when the distribution is such that $\bbeta$ is the true regression parameter,
        \item $\displaystyle\mathbb{E}_{\bbeta}\left[\left(Y-m(\bX, \bbeta)\right)^{2}
        \boldsymbol{h}\left(\bX,\bbeta\right)^{T}
        \boldsymbol{h}\left(\bX,\bbeta\right)\right]<\infty$,
        \item $\displaystyle\mathbb{E}_{\bbeta}\left[\left(Y-m(\bX, \bbeta)\right)^{2}
        \boldsymbol{h}\left(\bX,\bbeta\right)\boldsymbol{h}\left(\bX,\bbeta\right)^{T}\right]$ 
        is positive definite,
        \item $\displaystyle\mathbb{E}_{\bbeta}\left[\frac{\partial}{\partial\bbeta^{T}}
        \left(Y-m(\bX,\bbeta)\right)\boldsymbol{h}\left(\bX,\bbeta\right)\right]$ is non-singular,
        \item $\displaystyle\frac{1}{n_{\text{NP}}}\sum_{i\in S_{\text{NP}}}\frac{\partial}{\partial\bbeta^{T}}
        \left(Y_{i}-m(\bX_{i}, \bbeta)\right)\boldsymbol{h}\left(\bX_{i}, \bbeta\right)\longrightarrow\mathbb{E}_{\bbeta_{0}}\left[\frac{\partial}{\partial\bbeta^{T}}
        \left(Y-m(\bX, \bbeta)\right)\boldsymbol{h}\left(\bX, \bbeta\right)\right]$
        in probability, uniformly in $\bbeta$, when sampling is done from distribution where true regression parameter is $\bbeta_{0}$.
    \end{enumerate}
\end{itemize}

We do not however require all covariates which are significant for regression on $Y$ to be observed. Indeed consider the following toy model (the previously discussed regularity conditions are known to hold for linear regression if covariates are not too pathological):
\begin{align*}
    Y&=\beta_{0}+\beta_{1}X_{1}+\beta_{2}X_{2}+\epsilon\\
    \delta&=\frac{1}{1+\exp\left(-(X_{1}+\eta)\right)},
\end{align*}
where $\beta_{0}, \beta_{1}, \beta_{2}$, $\beta_{1}\neq0$ are constants, $X_1$ and $X_2$ are uncorrelated, random errors $\epsilon, \eta$ are independent and independent of $X_{2}$, then the estimators for $\beta_{0}', \beta_{1}$ in model:
\begin{align*}
    Y&=\beta_{0}'+\beta_{1}X_{1}+\epsilon',
\end{align*}
where $\beta_{0}'=\beta_{0}+\beta_{2}\mathbb{E}[X_{2}], \epsilon'=\epsilon+\beta_{2}(X_{2}-\mathbb{E}[X_{2}])$ are consistent since $Y$ and $\delta$ are independent given $X_{1}$ and assumptions for $\hat{y}-\hat{y}$ estimator are satisfied if the distribution of $X_{1}$ is not too pathological. If in addition $X_{1}$ has continuous distribution with support $\mathbb{R}$ and all $X_{1},X_{2},\epsilon$ have finite variance for all units in the population then assumptions for $\hat{y}-y$ are also satisfied. 

Notice that analogous model $Y=\beta_{0}''+\beta_{2}X_{2}+\epsilon''$ would not satisfy (A1.1).

\subsection{Sufficient conditions for assumption \eqref{linear-rep-assum}}\label{app-linear-rep}

Assumption \eqref{linear-rep-assum} requires that the estimator $\hat\bbeta$ has a linear representation of the form:
\begin{equation*}
    \hat\bbeta - \bbeta_0 = \frac{1}{n_{\text{NP}}} \sum_{j \in S_{\text{NP}}} \psi(Y_j, \bX_j) + o_P(n_{\text{NP}}^{-1/2})
\end{equation*}
where $\psi(\cdot)$ satisfies $\mathbb{E}[\psi(Y, \bX)] = 0$ and $\mathbb{E}[\|\psi(Y, \bX)\|^2] < \infty$. 

When $\hat\bbeta$ is defined as a solution to score equations from (A1.2), assumption \eqref{linear-rep-assum} follows directly from conditions (A1.2.1)--(A1.2.5) together with $\bbeta_0 \in \text{interior} (\mathcal{B})$, by \cite{Tsiatis2006}. The resulting influence function is:
\begin{equation*}
    \psi(Y, \bX) = -\bH^{-1}(Y-m(\bX, \bbeta_0)) \bh(\bX, \bbeta_0),
\end{equation*}
where $\bH = \mathbb{E}_{\bbeta_0} \left[ \dfrac{\partial}{\partial\bbeta^\top}(Y - m(\bX, \bbeta_0))\bh(\bX, \bbeta_0)\right]$ is the non-singular matrix from condition (A1.2.4).

In particular, \eqref{linear-rep-assum} is satisfied for the linear regression $m(\bx, \bbeta) = \bx^\top \bbeta$ with $\bh(\bx, \bbeta) = \bx$, conditions (A1.2.1)--(A1.2.5) reduce to nonsingularity of $\mathbb{E}[\bX\bX^\top]$ and the moment condition $\mathbb{E}[\|\bX\|^2 Y^2] < \infty$, and the influence function simplifies to:
\begin{equation*}
    \psi(Y, \bX) = \mathbb{E}[\bX\bX^\top]^{-1} \bX (Y-\bX^\top \bbeta_0).
\end{equation*}

\section{Main results}\label{app-main-results}

\subsection{Proof of Theorem \ref{yhat-y-consistency-proof}}\label{proof-of-yhat-y}
\begin{proof}
    The sufficiency of (\ref{conditional-mean-assum}) and \eqref{lim-res-prob-assum} is clear. Indeed, we can write:
    \begin{align}
        \label{haty-y-consistency-1}\frac{1}{N}\left|\sum_{i\in S_{\text{P}}}\frac{1}{\pi_{i}}
        \left(m\left(\bX_{i}, \hat{\bbeta}\right)-Y_{i}\right)\right|&\leq
        \frac{1}{N}\left|\sum_{i\in S_{\text{P}}}\frac{1}{\pi_{i}}
        \left(m\left(\bX_{i}, \hat{\bbeta}\right)-m\left(\bX_{i}, \bbeta_{0}\right)\right)\right|\\
        \label{haty-y-consistency-1.1}&+\frac{1}{N}\left|\sum_{i\in S_{\text{P}}}\frac{1}{\pi_{i}}
        \left(m\left(\bX_{i}, \bbeta_{0}\right)-Y_{i}\right)\right|.
    \end{align}
    By \eqref{lim-res-prob-assum}, expression in \eqref{haty-y-consistency-1.1} converges to $0$. The first term \eqref{haty-y-consistency-1} can be bounded as we assume that $m$ is Lipschitz, that is:
    \begin{equation}
        \left| m\left(\bX_i, \hat \bbeta \right) - m\left(\bX_i, \bbeta_0 \right) \right| \leq L(\bX_i) \| \hat \bbeta - \bbeta_0\|.
    \end{equation}
    Therefore, we have:
    \begin{equation*}
        \frac{1}{N}\sum_{i\in S_{\text{P}}}\frac{1}{\pi_{i}}
        \left| \left(m\left(\bX_{i}, \hat{\bbeta}\right)-m\left(\bX_{i}, \bbeta_{0}\right)\right)\right|
        \leq
        \left\|\hat\bbeta - \bbeta_0 \right\| \left( \frac{1}{N} \sum_{i \in S_{\text{P}}} \frac{1}{\pi_i} L(\bX_i) \right).
    \end{equation*}
    Notice that the second term is the HT-estimator of the population mean $\frac{1}{N}\sum_{i =1}^NL(\bX_i)$, and since $\mathbb{E}[L(\bX_i)] < \infty$, it follows that it is $O_P(1)$. Due to the fact that $\|\hat\bbeta - \bbeta_0\|$ converges to $0$ in probability and the second factor is $O_P(1)$, the expression above is $o_P(1)$. Combining these results, we get the required convergence of $\hat\mu_{MI} - \hat\mu_{HT}$ to zero in probability, and thus $\hat\mu_{MI} \rightarrow \hat\mu_{HT}$.
    
    Now to prove that the consistency of $\hat{\mu}_{\text{MI}}$ is sufficient, by the triangle inequality we have:
    \begin{align}\label{haty-y-consistency-3}
        &\frac{1}{N}\left|\sum_{i\in S_{\text{P}}}\frac{1}{\pi_{i}}
        \frac{1}{k}\sum_{t=1}^{k}Y_{\hat{\nu}(i, t)}-
        \sum_{i\in S_{\text{P}}}\frac{1}{\pi_{i}}Y_{i}\right|\leq\\
        &\frac{1}{N}\left|\sum_{i\in S_{\text{P}}}\frac{1}{\pi_{i}}\left(
        \frac{1}{k}\sum_{t=1}^{k}Y_{\hat{\nu}(i, t)}-
        m\left(\bX_{i}, \hat{\bbeta}\right)\right)\right|+
        \underbrace{\frac{1}{N}\left|\sum_{i\in S_{\text{P}}}\frac{1}{\pi_{i}}
        \left(m\left(\bX_{i}, \hat{\bbeta}\right)-Y_{i}\right)\right|}_{\displaystyle\overset{n_{\text{NP}}\rightarrow\infty}{\longrightarrow}0
        \text{ in probability}},\nonumber
    \end{align}
    where the second term converges to zero by the assumption of consistency of $\hat\mu_{MI}$. Notice that from the continuity of $m$, we have the following implications:
    \begin{align}\label{haty-y-consistency-continuity}
        &\forall \varepsilon'>0:\mathbb{P}\left[\left.
        |\bx_{i}-\bX|<\varepsilon'\right|\{i\in S_{\text{P}}\}\right]>0,\\
        \implies&\forall \varepsilon>0,\bbeta:
        \mathbb{P}\left[\left.| m(\bx_{i}, \bbeta)-
        m(\bX, \bbeta)|<\varepsilon\right|\{i\in S_{\text{P}}\}\right]>0.\nonumber
    \end{align}
    \noindent for $\bbeta$ in some neighbourhood of $\bbeta_{0}$. Therefore, by \eqref{yhat-y-assum} with positive probability, we have:
    \begin{equation*}
        \forall\varepsilon>0, \bbeta:
        \{\exists j\in S_{\text{NP}}:|Y_{j}-m(\bX_{i}, \bbeta)|<
        \varepsilon\}\cap\{i\in S_{\text{P}}\}.
    \end{equation*}
    \noindent Notice that for $\hat{y}-y$ matching, $\displaystyle \frac{1}{k}\sum_{t=1}^{k}\left(y_{\hat{\nu}(i, t)}-m\left(\bx_{i}, \hat{\bbeta}\right)\right)$ is exactly the minimized quantity from Algorithm \ref{algo-1}. By virtue of the above reasoning and assumption (\ref{asymptotic-neighbour-assum}), we have:
    \begin{align}\label{haty-y-consistency-5}
        \forall \varepsilon>0, i:
        &\lim_{n_{\text{NP}}\rightarrow\infty}
        \mathbb{P}\left[\left.k\leq\#\left\{j\in S_{\text{NP}}:
        \left|Y_{j}-m\left(\bX_{i}, \hat{\bbeta}\right)\right|<
        \varepsilon\right\}\right|\{i\in S_{\text{P}}\}\right]=1.
    \end{align}
    Next, we can bound the first term in \eqref{haty-y-consistency-3} as:
    \begin{align*}
        \frac{1}{N}\left|\sum_{i\in S_{\text{P}}}\frac{1}{\pi_{i}}\left(
        \frac{1}{k}\sum_{t=1}^{k}Y_{\hat{\nu}(i, t)}-
        m\left(\bX_{i}, \hat{\bbeta}\right)\right)\right| \leq \frac{1}{N}\sum_{i\in S_{\text{P}}}\frac{1}{\pi_{i}}
        \underbrace{\left|\frac{1}{k}\sum_{t=1}^{k}Y_{\hat{\nu}(i, t)}-
        m\left(\bX_{i}, \hat{\bbeta}\right)\right|}_{:= \Delta_i}.
    \end{align*}
    Using the result in \eqref{haty-y-consistency-5} we have that: \begin{equation*}
        \forall \varepsilon > 0, i:\lim_{n_{\text{NP}} \rightarrow \infty} \mathbb{P}[ \Delta_i> \varepsilon | \{i \in S_{\text{P}}\}] = 0
    \end{equation*}
    since asymptotically there is at least $k$ donors in $S_{\text{NP}}$ with $|Y_i - m(\bX_i, \hat\bbeta)| < \varepsilon$. Hence, $\Delta_i \rightarrow 0$ in probability. Moreover, as $Y$ has finite second moment, we can conclude that: $\mathbb{E}\left[\left(\frac{1}{k} \sum_{t=1}^k Y_{\hat\nu(i,t)}\right)^2\right] < \infty$, and as $|m(\bX_i, \hat\bbeta)| \leq m(\bX_i, \bbeta_0) + L(\bX_i)\|\hat\bbeta - \bbeta_0\|$, by assumptions \eqref{conditional-mean-assum}, \eqref{lipschitz-assum}, we also have that: $\mathbb{E}[m(\bX_i, \hat\bbeta)^2] < \infty$. Consequently: $\sup \mathbb{E}[\Delta_i^2] < \infty$, which implies that ${\Delta_i}$ is uniformly integrable. Then, since also $\Delta_i \overset{P}{\rightarrow} 0$, we have that: $\mathbb{E}[\Delta_i] \rightarrow 0$. Next, notice that:
    \begin{equation*}
        \mathbb{E}\left[ \frac{1}{N}\sum_{i\in S_{\text{P}}}\frac{1}{\pi_{i}}
        \Delta_i \right] = \mathbb{E}\left[ \frac{1}{N}\sum_{i=1}^N \Delta_i \right]
    \end{equation*}
    and this expression converges to zero. Now, since $\frac{1}{N}\sum_{i\in S_{\text{P}}}\frac{1}{\pi_{i}} \Delta_i \geq 0$ using the Markov inequality we obtain:
    \begin{equation*}
        \mathbb{P}\left[ \frac{1}{N}\sum_{i\in S_{\text{P}}}\frac{1}{\pi_{i}}
        \Delta_i > \varepsilon \right] \leq \frac{1}{\varepsilon} \mathbb{E}\left[ \frac{1}{N}\sum_{i\in S_{\text{P}}}\frac{1}{\pi_{i}}
        \Delta_i \right] \rightarrow 0
    \end{equation*}
    so $ \frac{1}{N}\sum_{i\in S_{\text{P}}}\frac{1}{\pi_{i}}\Delta_i \rightarrow 0$ in probability. Consequently, the first term in \eqref{haty-y-consistency-3} converges to zero in probability. Thus, the $\hat{y}-y$ matching estimator from Algorithm \ref{algo-1} is consistent.     
    To prove that the consistency of $\hat{\mu}_{\text{MI}}$ is necessary see that $ \frac{1}{N}\sum_{i\in S_{\text{P}}}\frac{1}{\pi_{i}}\Delta_i \overset{P}{\rightarrow} 0$ still stands and that if $\hat{y}-y$ estimator is consistent we have:
    \begin{align*}
        &\frac{1}{N}\left|\sum_{i\in S_{\text{P}}}\frac{1}{\pi_{i}}
        \left(m\left(\bX_{i}, \hat{\bbeta}\right)-Y_{i}\right)\right|\leq\\
        &\frac{1}{N}\left|\sum_{i\in S_{\text{P}}}\frac{1}{\pi_{i}}\left(
        \frac{1}{k}\sum_{t=1}^{k}Y_{\hat{\nu}(i, t)}-
        m\left(\bX_{i}, \hat{\bbeta}\right)\right)\right|+
        \frac{1}{N}\left|\sum_{i\in S_{\text{P}}}\frac{1}{\pi_{i}}
        \frac{1}{k}\sum_{t=1}^{k}Y_{\hat{\nu}(i, t)}-
        \sum_{i\in S_{\text{P}}}\frac{1}{\pi_{i}}Y_{i}\right|,
    \end{align*}
    which converges to $0$.
\end{proof}

\subsection{Proof of Theorem \ref{yhat-yhat-consistency-proof}}\label{proof-of-yhat-yhat}

\begin{proof} The objective expression for proving the sufficiency of consistency of $\hat{\mu}_{\text{MI}}$:
\begin{equation*}
    \frac{1}{N}\left|\sum_{i\in S_{\text{P}}}\frac{1}{\pi_{i}}
    \frac{1}{k}\sum_{t=1}^{k}Y_{\hat{\nu}(i, t)}-
    \sum_{i\in S_{\text{P}}}\frac{1}{\pi_{i}}Y_{i}\right|,
\end{equation*}
can be subdivided into three parts:
\begin{align}
    &\frac{1}{N}\left|\sum_{i\in S_{\text{P}}}\frac{1}{\pi_{i}}
    \frac{1}{k}\sum_{t=1}^{k}\left(Y_{\hat{\nu}(i, t)}-Y_{i}\right)\right|\nonumber\leq\\
    \label{yhat-yhat-division-1}&\frac{1}{N}\sum_{i\in S_{\text{P}}}\frac{1}{\pi_{i}}
    \left|\frac{1}{k}\sum_{t=1}^{k}m\left(\bX_{i}, \hat{\bbeta}\right)-
    m\left(\bX_{\hat{\nu}(i, t)}, \hat{\bbeta}\right)\right|+\\
    \label{yhat-yhat-division-2}&\frac{1}{N}\left|\sum_{i\in S_{\text{P}}}\frac{1}{\pi_{i}}
    \left(Y_{i}-m\left(\bX_{i}, \hat{\bbeta}\right)\right)\right|+\\
    \label{yhat-yhat-division-3}&\frac{1}{N}\left|\sum_{i\in S_{\text{P}}}\frac{1}{\pi_{i}}
    \frac{1}{k}\sum_{t=1}^{k}
    \left(Y_{\hat{\nu}(i, t)}-m\left(\bX_{\hat{\nu}(i, t)}, \hat{\bbeta}\right)\right)\right|,
\end{align}
\noindent where \eqref{yhat-yhat-division-2} tends to $0$ by assumption. By (\ref{asymptotic-neighbour-assum}), \eqref{yhat-yhat-assum}, and \eqref{haty-y-consistency-continuity} we have:
\begin{equation}\label{yhat-yhat-consistency-4}
    \forall \varepsilon>0, i:
    \lim_{n_{\text{NP}}\rightarrow\infty}
    \mathbb{P}\left[\left.k\leq\#\left\{j\in S_{\text{NP}}:
    \left|m\left(\bX_{i}, \hat{\bbeta}\right)-
    m\left(\bX_{j}, \hat{\bbeta}\right)\right|<
    \varepsilon\right\}\right|\{i\in S_{\text{P}}\}\right]=1.
\end{equation}
\noindent In other words, asymptotically there will be at least $k$ units in $S_{\text{NP}}$ that are "similar" to $i\in S_{\text{P}}$, and we can asymptotically replace $m(\bx_{i},\hat{\bbeta})$ with the $m(\bx,\hat{\bbeta})$ value of any of those $k$ units. Let $\Delta^A_i=\left|\frac{1}{k}\sum_{t=1}^{k}\left(m(\bX_{i}, \hat{\bbeta})-m(\bX_{\hat{\nu}(i, t)}, \hat{\bbeta})\right)\right|$. Using the analogous approach as in the proof of Theorem \ref{yhat-y-consistency-proof} and the expression in \eqref{yhat-yhat-consistency-4}, we have that
\begin{equation*}
    \mathbb{E}\left[\frac{1}{N}\sum_{i\in S_{\text{P}}}\frac{1}{\pi_{i}} \Delta^A_i \right] \overset{P}{\rightarrow} 0.
\end{equation*}
Therefore, since \eqref{yhat-yhat-division-1} is non-negative, we can use the Markov inequality to show that:
\begin{equation*}
    \frac{1}{N}\sum_{i\in S_{\text{P}}}\frac{1}{\pi_{i}} \Delta^A_i \overset{P}{\rightarrow} 0,
\end{equation*}
i.e. \eqref{yhat-yhat-division-1} converges to zero in probability.

Now for \eqref{yhat-yhat-division-3}, we have:
\begin{align*}
    Y_{\hat{\nu}(i, t)}-m\left(\bX_{\hat{\nu}(i, t)}, \hat{\bbeta}\right) &= \underbrace{Y_{\hat{\nu}(i, t)}-m\left(\bX_{\hat{\nu}(i, t)}, \bbeta_0\right)}_{= \varepsilon_{\hat{\nu}(i, t)}} + \left(m\left(\bX_{\hat{\nu}(i, t)}, \bbeta_0\right)\ - m\left(\bX_{\hat{\nu}(i, t)}, \hat{\bbeta}\right)\right) 
\end{align*}
where the second term can be bounded using assumption \eqref{lipschitz-assum} and thus we focus on the $\varepsilon_{\hat{\nu}(i, t)}$. 
We can write:
\begin{align}
     \frac{1}{N}\sum_{i \in S_{\text{P}}} \frac{1}{\pi_i} \frac{1}{k} \sum_{t=1}^k \varepsilon_{\hat{\nu}(i, t)}
    &= \sum_{j \in S_{\text{NP}}}  \frac{1}{N} \sum_{i \in S_{\text{P}}} \frac{1}{\pi_i} \frac{1}{k} \sum_{t=1}^k \varepsilon_j \1_{\{j = {\hat{\nu}(i, t)}\}} =\nonumber\\
    &= \sum_{j \in S_{\text{NP}}} \varepsilon_j \left(\frac{1}{N} \sum_{i \in S_{\text{P}}} \frac{1}{\pi_i} \frac{1}{k} \sum_{t=1}^k \1_{\{j = {\hat{\nu}(i, t)}\}}\right) = \nonumber\\
    \label{yhat-yhat-proof-weights}&= \sum_{j \in S_{\text{NP}}} \varepsilon_j W_{j, N},
\end{align}
where $W_{j,N}$ is exactly the same value as defined in assumption \eqref{donor-reuse-assum}. Using the law of total variance, we can write it as a sum of two components:
\begin{equation*}
    \mathbb{V} \left[ \sum_{j \in S_{\text{NP}}} \varepsilon_j W_{j, N}\right] = \underbrace{\mathbb{E} \left[\mathbb{V}\left[\sum_{j \in S_{\text{NP}}} \varepsilon_j W_{j, N} \mid \bX, \hat\bbeta\right] \right]}_* +  \underbrace{\mathbb{V} \left[\mathbb{E}\left[\sum_{j \in S_{\text{NP}}} \varepsilon_j W_{j, N} \mid \bX, \hat\bbeta\right] \right]}_{**}.
\end{equation*}
Starting with $*$, note that $W_{j, N}$ depends on $\hat\bbeta$ and on $\bX_i$ for all $i \in S_{\text{P}}$ and $\bX_j$ for all $j \in S_{\text{NP}}$. Conditionally on $\bX, \hat\bbeta$, the weights $W_{j, N}$ are therefore deterministic and we get:
\begin{equation*}
    \mathbb{V}\left[\sum_{j \in S_{\text{NP}}} \varepsilon_j W_{j, N} \mid \bX, \hat\bbeta\right] = \sum_{j \in S_{\text{NP}}} W_{j, N}^2 \cdot \mathbb{V}\left[\varepsilon_j \mid \bX, \hat\bbeta\right] +  \sum_{\substack{j,l \in S_{\text{NP}}\\j\neq l}} W_{j, N} \cdot W_{l, N} \cdot \text{cov}\left[\varepsilon_j, \varepsilon_l \mid \bX, \hat\bbeta\right].
\end{equation*}
Then: 
\begin{equation*}
    \sum_{j \in S_{\text{NP}}} W_{j, N}^2 \cdot \mathbb{V}\left[\varepsilon_j \mid \bX, \hat\bbeta\right] \leq C \cdot \sum_{j \in S_{\text{NP}}} W_{j, N}^2 \overset{P}{\rightarrow} 0
\end{equation*}
by the assumptions of finite moments of the error term and \eqref{donor-reuse-assum}, and:
\begin{align*}
    \sum_{\substack{j,l \in S_{\text{NP}}\\j\neq l}} W_{j, N} \cdot W_{l, N} \cdot \text{cov}\left[\varepsilon_j, \varepsilon_l \mid \bX, \hat\bbeta\right] &\leq \sum_{\substack{j,l \in S_{\text{NP}}\\j\neq l}} |W_{j, N}| \cdot |W_{l, N}| \cdot |\text{cov}\left[\varepsilon_j, \varepsilon_l \mid \bX, \hat\bbeta\right]| \leq \\
    & \leq O_P(n_{\text{NP}}^{-1}) \sum_{\substack{j,l \in S_{\text{NP}}\\j\neq l}} |W_{j, N}| \cdot |W_{l, N}| \leq \\
    & \leq O_P(n_{\text{NP}}^{-1}) \underbrace{\left(\sum_{j \in S_{\text{NP}}} W_{j, N} \right)^2}_{O_P(1)}  = O_P(n_{\text{NP}}^{-1})
\end{align*}
using the property $|\text{cov}\left[\varepsilon_j, \varepsilon_l \mid \bX_j, \hat\bbeta\right]| = O_P(n_{\text{NP}}^{-1})$ and the fact:
\begin{equation}\label{sum-wjn-op1}
    \sum_{j \in S_{\text{NP}}} W_{j, N} = \frac{1}{N} \sum_{i \in S_{\text{P}}} \frac{1}{\pi_i}\frac{1}{k }\sum_{t =1}^k \underbrace{\sum_{j \in S_{\text{NP}}} \1 \{\hat{\nu}(i, t) = j \}}_{=1} = \frac{1}{N} \sum_{i \in S_{\text{P}}} \frac{1}{\pi_i} \overset{\eqref{pop-sequence-assum}}{\leq} C_2 < \infty.
\end{equation}
Therefore we have:
\begin{equation*}
    \mathbb{V}\left[\sum_{j \in S_{\text{NP}}} \varepsilon_j W_{j, N} \mid \bX, \hat\bbeta\right] \overset{P}{\rightarrow} 0.
\end{equation*}
Next, we focus on $**$, where $W_{j,N}$ becomes deterministic conditionally on $\bX, \hat\bbeta$:
\begin{equation*}
    \mathbb{E}\left[\sum_{j \in S_{\text{NP}}} \varepsilon_j W_{j, N} \mid \bX, \hat\bbeta\right] = \sum_{j \in S_{\text{NP}}} W_{j,N} \cdot \mathbb{E}\left[\varepsilon_j\mid \bX, \hat\bbeta\right].
\end{equation*}
Using \eqref{conditional-mean-assum} we have that:
\begin{equation*}
    \mathbb{E}\left[\varepsilon_j\mid \bX, \hat\bbeta\right] = m(\bX_j, \hat\bbeta) - m(\bX_j, \bbeta_0) + O_P(n_{\text{NP}}^{-1})
\end{equation*}
where $O_P(n_{\text{NP}}^{-1})$ accounts for the dependence of $\hat\bbeta$ on $Y_j, j \in S_{\text{NP}}$ by the assumption \eqref{linear-rep-assum}. Then, using the assumption \eqref{lipschitz-assum}:
\begin{equation*}
    |m(\bX_j, \hat\bbeta) - m(\bX_j, \bbeta_0)| \leq L(\bX_j) \|\hat\bbeta - \bbeta_0\|
\end{equation*}
and we get:
\begin{align*}
    \mathbb{E}\left[\sum_{j \in S_{\text{NP}}} \varepsilon_j W_{j, N} \mid \bX, \hat\bbeta\right] & \leq \left|\sum_{j \in S_{\text{NP}}} W_{j,N} \cdot \mathbb{E}\left[\varepsilon_j\mid \bX, \hat\bbeta\right] \right|  \leq \\
    & \leq \left|\sum_{j \in S_{\text{NP}}} W_{j,N} \cdot  L(\bX_j) \|\hat\bbeta - \bbeta_0\| + O_P(n_{\text{NP}}^{-1}) \sum_{j \in S_{\text{NP}}} W_{j,N}\right| = \\
    & = \|\hat\bbeta - \bbeta_0\|\sum_{j \in S_{\text{NP}}} W_{j,N} \cdot  L(\bX_j)  + O_P(n_{\text{NP}}^{-1}) \sum_{j \in S_{\text{NP}}} W_{j,N}.
\end{align*}
Notice that:
\begin{equation*}
    \sum_{j \in S_{\text{NP}}} W_{j,N} \cdot  L(\bX_j) = \frac{1}{N} \sum_{i \in S_{\text{P}}} \frac{1}{\pi_i} \frac{1}{k} \sum_{t=1}^k L(\bX_{\hat{\nu}(i,t)}).
\end{equation*}
Then:
\begin{equation*}
    \mathbb{E}\left[\sum_{j \in S_{\text{NP}}} W_{j,N} \cdot  L(\bX_j)\right] = \frac{1}{N} \sum_{i \in S_{\text{P}}} \frac{1}{\pi_i} \frac{1}{k} \sum_{t=1}^k \mathbb{E}\left[ L(\bX_{\hat\nu(i,t)}) \right] \leq \frac{1}{N} \sum_{i \in S_{\text{P}}} \frac{1}{\pi_i} \cdot C \leq \frac{C}{C_1} < \infty
\end{equation*}
and by the Markov inequality: $\sum_{j \in S_{\text{NP}}} W_{j,N} \cdot  L(\bX_j) = O_P(1)$. Therefore, using the result in \eqref{sum-wjn-op1}, we get:
\begin{equation*}
    \mathbb{E}\left[\sum_{j \in S_{\text{NP}}} \varepsilon_j W_{j, N} \mid \bX, \hat\bbeta\right] = \underbrace{\|\hat\bbeta - \bbeta_0\|}_{O_P(n_{\text{NP}}^{-1/2})} \underbrace{\sum_{j \in S_{\text{NP}}} W_{j,N} \cdot  L(\bX_j)}_{O_P(1)}  + O_P(n_{\text{NP}}^{-1}) \underbrace{\sum_{j \in S_{\text{NP}}} W_{j,N}}_{O_P(1)} = O_P(n_{\text{NP}}^{-1/2}).
\end{equation*}
Since the above expression is $O_P(n_{\text{NP}}^{-1/2})$, its variance satisfies $\mathbb{V}[\cdot] \leq \mathbb{E}[O_P(n_{\text{NP}}^{-1})]$. By the boundedness of $\sum_j W_{j,N} = O_P(1)$ and finite second moments of $\varepsilon_j$, the $O_P(n_{\text{NP}}^{-1})$ term has finite expectation, so $\mathbb{V}[\cdot] = O(n_{\text{NP}}^{-1}) \rightarrow 0$.
Decomposing the expression into two parts, we showed that both of them converge to $0$. By the Chebyshev inequality, we get that for any $\delta >0$:
\begin{equation}
    \mathbb{P}\left(\left| \sum_{j \in S_{\text{NP}}} \varepsilon_j W_{j, N}  \right|>\delta\right) \leq \frac{\mathbb{V}\left[ \sum_{j \in S_{\text{NP}}} \varepsilon_j W_{j, N}  \right]}{\delta^2} \rightarrow 0
\end{equation}
and therefore, the term in \eqref{yhat-yhat-division-3} also converges to $0$, which was the last needed step to show the consistency of the estimator in \eqref{mu-hat-definition-yhat-yhat-match}.

\end{proof}

\subsection{Proof of Theorem \ref{yhat-yhat-consistency-mis-specified}}\label{proof-mis-specified}

    \begin{proof}
    Set $\varepsilon>0$ and recall from \eqref{yhat-yhat-consistency-4} that for each $i\in S_{\text{P}}$ with probability one asymptotically we will have at least $k$ units in $S_{\text{NP}}$ such that:
    \begin{equation*}
        \left|m\left(\bX_{i}, \hat{\bbeta}\right)-m\left(\bX_{j}, \hat{\bbeta}\right)\right|<\varepsilon,
    \end{equation*}
    but since:
    \begin{align*}
        &\left|m\left(\bX_{i}, \bbeta^{\ast}\right)-m\left(\bX_{j}, \bbeta^{\ast}\right)\right|\leq
        \left|m\left(\bX_{i}, \hat{\bbeta}\right) -m\left(\bX_{j},  \hat{\bbeta}\right)\right|+\\
        &\left|m\left(\bX_{j}, \hat{\bbeta}\right) -m\left(\bX_{j}, \bbeta^{\ast}\right)\right|
        +\left|m\left(\bX_{i}, \hat{\bbeta}\right) -m\left(\bX_{i}, \bbeta^{\ast}\right)\right|,
    \end{align*}
    we have that these same $k$ units in $S_{\text{NP}}$ will satisfy:
    \begin{equation*}
        \left|m\left(\bX_{i}, \bbeta^{\ast}\right)-m\left(\bX_{j}, \bbeta^{\ast}\right)\right|<3\varepsilon.
    \end{equation*}
    We can write:
    \begin{align}
        \label{llllll=1}\frac{1}{N}\left|\sum_{i\in S_{\text{P}}}\frac{1}{k\pi_{i}}
        \sum_{t=1}^{k}(Y_{i}-Y_{\hat{\nu}(i,t)})\right|&\leq
        \underbrace{\frac{1}{N}\left|\sum_{i\in S_{\text{P}}}\frac{1}{\pi_{i}}
        (Y_{i}-m'(\bX_{i}, \bbeta_{0}))\right|}_{\rightarrow0 \text{ in probability by \eqref{lim-res-prob-assum}}}+\\
        \label{misspec-division-2}&\frac{1}{N}\left|\sum_{i\in S_{\text{P}}}\frac{1}{k\pi_{i}}
        \sum_{t=1}^{k}(Y_{\hat{\nu}(i,t)}-m'(\bX_{\hat{\nu}(i, t)}, \bbeta_{0}))\right|+\\
        \label{misspec-division-3}&\frac{1}{N}\left|\sum_{i\in S_{\text{P}}}\frac{1}{k\pi_{i}}
        \sum_{t=1}^{k}(m'(\bX_{i}, \bbeta_{0})-m'(\bX_{\hat{\nu}(i, t)}, \bbeta_{0}))\right|.
    \end{align}
    
    First, notice that we can rewrite \eqref{misspec-division-2} as:
    \begin{equation*}
        \frac{1}{N}\sum_{i\in S_{\text{P}}}\frac{1}{k\pi_{i}}
        \sum_{t=1}^{k}\underbrace{(Y_{\hat{\nu}(i,t)}-m'(\bX_{\hat{\nu}(i, t)}, \bbeta_{0}))}_{\varepsilon'_{\hat{\nu}(i,t)}} = \frac{1}{N}\sum_{i\in S_{\text{P}}} \frac{1}{k\pi_{i}} \sum_{t=1}^{k}\varepsilon'_{\hat{\nu}(i,t)} = \sum_{j \in S_{\text{NP}}} \varepsilon'_j W_{j,N}
    \end{equation*}
    where $\varepsilon_j' = Y_j - m'(\bX_j, \bbeta_{0})$ satisfies $\mathbb{E}[\varepsilon_j' \mid \bX_j] = 0$ and $\mathbb{E}[\varepsilon_j^{'2}] < \infty$ by assumption \eqref{conditional-mean-assum} for $m'$ and the fact that $\mathbb{E}[Y_j^2] < \infty$. The weights $W_{j,N}$ are identical to those in \eqref{yhat-yhat-proof-weights}. Therefore, the argument in the proof of Theorem~\ref{yhat-yhat-consistency-proof} can be applied, with $\varepsilon_j'$ instead of $\varepsilon_j$, and assumption \eqref{linear-rep-assum} used in the same way. Hence, \eqref{misspec-division-2} converges to~$0$ in probability.

    Now we show that \eqref{misspec-division-3} converges to $0$ in $L^{1}$ and consequently in probability. Since $\mathbb{E}[Y|\bX] = m'(\bX, \bbeta_0)$, we can write:
    \begin{align*}
        \left|m'(\bX_{i}, \bbeta_{0})-
        m'(\bX_{\hat{\nu}(i, t)}, \bbeta_{0})\right| &= |\mathbb{E}[Y_i - Y_{\hat{\nu}(i, t)} \mid \bX_i, \bX_{\hat{\nu}(i, t)}]| \leq \\
        & \leq \mathbb{E}[|Y_i - Y_{\hat{\nu}(i, t)}| \mid \bX_i, \bX_{\hat{\nu}(i, t)}]
    \end{align*}
    by the Jensen's inequality. Taking the conditional expectation given $m(\bX_i, \bbeta^*), m(\bX_{\hat{\nu}(i, t)}, \bbeta^*)$ and using the assumption \eqref{miss-exp-assum}:
    \begin{equation*}
        \mathbb{E}[|Y_i - Y_{\hat{\nu}(i, t)}| \mid m(\bX_{i}, \bbeta^*), m(\bX_{\hat{\nu}(i, t)}, \bbeta^*)]  \leq C |m(\bX_{i}, \bbeta^*) - m(\bX_{\hat{\nu}(i, t)}, \bbeta^*)|.
    \end{equation*}
    By the argument at the beginning of the proof, for each $\varepsilon > 0$ the matched units satisfy $\left|m\left(\bX_{i}, \bbeta^{\ast}\right)-m\left(\bX_{j}, \bbeta^{\ast}\right)\right|<3\varepsilon$ with probability tending to $1$. Therefore:
    \begin{equation*}
        |m'(\bX_{i}, \bbeta_0) - m'(\bX_{\hat{\nu}(i, t)}, \bbeta_0)| \overset{P}{\rightarrow} 0.
    \end{equation*}
    The last step is to exchange the limit and expectation through uniform integrability. By Jensen's inequality and the assumption \eqref{miss-exp-assum}:
    \begin{equation*}
        |m'(\bX_{i}, \bbeta_0) - m'(\bX_{\hat{\nu}(i, t)}, \bbeta_0)| \leq \mathbb{E}[|Y_i - Y_{\hat{\nu}(i, t)}| \mid \bX_i, \bX_{\hat{\nu}(i, t)}].
    \end{equation*}
    Squaring both sides and taking the expectation:
    \begin{equation*}
        \mathbb{E}\left[|m'(\bX_i, \bbeta_0) - m'(\bX_{\hat\nu(i, t)}, \bbeta_0)|^2 \right] \leq C^2 \mathbb{E}\left[|m(\bX_i, \bbeta^*) - m(\bX_{\hat\nu(i, t)}, \bbeta^*)|^2 \right].
    \end{equation*}
    The right side of this expression is bounded uniformly in $i$ and $t$. The reasoning is that since $\mathbb{E}[Y^2] < \infty$, assumption \eqref{conditional-mean-assum} for $m'$ and assumption \eqref{miss-limit-assum} together imply that $m(\bX, \bbeta^*)$ has finite second moment, so:
    \begin{equation*}
        \mathbb{E}[|m(\bX_i, \bbeta^*)|^2] \leq 4 \mathbb{E}[m(\bX, \bbeta^*)^2] < \infty.
    \end{equation*}
    Therefore $\{|m'(\bX_i, \bbeta_0) - m'(\bX_j, \bbeta_0)|\}$ is uniformly integrable (by the $L^2$ boundedness criterion) and combined with the convergence in probability shown above, it implies the convergence in $L^1$:
    \begin{equation*}
         \mathbb{E}\left[|m'(\bX_i, \bbeta_0) - m'(\bX_{\hat\nu(i, t)}, \bbeta_0)| \right] \rightarrow 0.
    \end{equation*}
    Consequently, by assumption \eqref{pop-sequence-assum}
    \begin{align*}
        \mathbb{E}\left[\frac{1}{N} \left|\sum_{i \in S_{\text{P}}} \frac{1}{\pi_i} \frac{1}{k} \sum_{t=1}^k (m'(\bX_i, \bbeta_0) - m'(\bX_{\hat\nu(i,t)}, \bbeta_0))\right|\right]& \\\leq \frac{1}{C_1 k} \sum_{t=1}^k \mathbb{E}\left[\left|m'(\bX_i, \bbeta_0) - m'(\bX_{\hat\nu(i,t)}, \bbeta_0)\right|\right] &\rightarrow 0
    \end{align*}
    and thus \eqref{misspec-division-3} converges to $0$ in probability by the Markov's inequality.

\end{proof}

\subsection{Proof of Theorem \ref{variance-pmm-proof}}\label{proof-of-2}

\begin{proof}
Using law of total variance we have:
\begin{align}
    \mathbb{V}\left[\hat{\mu}\right]&=
    \frac{1}{N^{2}k^{2}}
    \mathbb{E}\left[\mathbb{V}\left[\left.
    \sum_{i \in S_{\text{P}}}
    \frac{1}{\pi_{i}}\sum_{t=1}^{k}Y_{\hat{\nu}(i, t)}\right|
    \sigma\left(\{Y_{i}, \bX_{i}, \hat{\nu}(i, t), \pi_{i}\}_{i\in U, t\in\{1,\dots,k\}}\right)\right]\right]
    \label{Law-total-variance-1-yhat-y}\\
    &+\frac{1}{N^{2}k^{2}}
    \mathbb{V}\left[\mathbb{E}\left[\left.
    \sum_{i \in S_{\text{P}}}
    \frac{1}{\pi_{i}}\sum_{t=1}^{k}Y_{\hat{\nu}(i, t)}\right|
    \sigma\left(\{Y_{i}, \bX_{i}, \hat{\nu}(i, t), \pi_{i}\}_{i\in U, t\in\{1,\dots,k\}}\right)\right]\right].
    \label{Law-total-variance-2-yhat-y}
\end{align}

by Cauchy product \eqref{Law-total-variance-2-yhat-y} can be expressed as:
\begin{align*}
    &\frac{1}{N^{2}k^{2}}
    \mathbb{V}\left[\mathbb{E}\left[\left.
    \sum_{i=1}^{N}\frac{1}{\pi_{i}}\1_{\{i\in S_{\text{P}}\}}\sum_{t=1}^{k}Y_{\hat{\nu}(i, t)}
    \right|\sigma\left(\{Y_{i}, \bX_{i}, \hat{\nu}(i, t), \pi_{i}\}_{i\in U, t\in\{1,\dots,k\}}\right)\right]\right]\\
    &=\frac{1}{N^{2}k^{2}}
    \mathbb{V}\left[\sum_{i=1}^{N}\sum_{t=1}^{k}Y_{\hat{\nu}(i, t)}\right]
    =\frac{1}{N^{2}k^{2}}
    \mathbb{E}\left[\left(\sum_{i=1}^{N}\sum_{t=1}^{k}Y_{\hat{\nu}(i, t)}\right)^{2}\right]-
    \frac{1}{N^{2}k^{2}}
    \left(\mathbb{E}\left[\sum_{i=1}^{N}\sum_{t=1}^{k}Y_{\hat{\nu}(i, t)}\right]\right)^{2}\\
    &=\frac{1}{N^{2}k^{2}}
    \mathbb{E}\left[\left(\sum_{i=1}^{N}\sum_{t=1}^{k}Y_{\hat{\nu}(i, t)}\right)
    \left(\sum_{j=1}^{N}\sum_{t'=1}^{k}Y_{\hat{\nu}(j, t')}\right)\right]\\
    &-\frac{1}{N^{2}k^{2}}
    \left(\sum_{i=1}^{N}\mathbb{E}
    \left[\sum_{t=1}^{k}Y_{\hat{\nu}(i, t)}\right]\right)
    \left(\sum_{j=1}^{N}\mathbb{E}
    \left[\sum_{t'=1}^{k}Y_{\hat{\nu}(j, t')}\right]\right)\\
    &=\frac{1}{N^{2}k^{2}}
    \mathbb{E}\left[\sum_{i=1}^{N}\sum_{j=1}^{N}
    \sum_{t'=1}^{k}\sum_{t=1}^{k}
    Y_{\hat{\nu}(i, t)}Y_{\hat{\nu}(j, t')}\right]\\
    &-\frac{1}{N^{2}k^{2}}
    \left(\sum_{i=1}^{N}\sum_{j=1}^{N}\mathbb{E}
    \left[\sum_{t=1}^{k}Y_{\hat{\nu}(i, t)}\right]
    \mathbb{E}
    \left[\sum_{t'=1}^{k}Y_{\hat{\nu}(j, t')}\right]\right)\\
    &=\frac{1}{N^{2}k^{2}}
    \sum_{i=1}^{N}\sum_{j=1}^{N}
    \sum_{t'=1}^{k}\sum_{t=1}^{k}
    \mathbb{E}\left[Y_{\hat{\nu}(i, t)}Y_{\hat{\nu}(j, t')}\right]\\
    &-\frac{1}{N^{2}k^{2}}
    \left(\sum_{i=1}^{N}\sum_{j=1}^{N}
    \sum_{t=1}^{k}\sum_{t'=1}^{k}
    \mathbb{E}\left[Y_{\hat{\nu}(i, t)}\right]
    \mathbb{E}\left[Y_{\hat{\nu}(j, t')}\right]\right)\\
    &=\frac{1}{N^{2}k^{2}}
    \sum_{i=1}^{N}\sum_{j=1}^{N}
    \sum_{t'=1}^{k}\sum_{t=1}^{k}
    \left(\mathbb{E}\left[Y_{\hat{\nu}(i, t)}Y_{\hat{\nu}(j, t')}\right]-
    \mathbb{E}\left[Y_{\hat{\nu}(i, t)}\right]
    \mathbb{E}\left[Y_{\hat{\nu}(j, t')}\right]\right)\\
    &=\frac{1}{N^{2}}
    \sum_{i=1}^{N}\sum_{j=1}^{N}
    \text{cov}\left(
    \frac{1}{k}\sum_{t=1}^{k}Y_{\hat{\nu}(i, t)},
    \frac{1}{k}\sum_{t'=1}^{k}Y_{\hat{\nu}(j, t')}\right),
\end{align*}

\noindent which can be estimated by:
\begin{equation}\label{yhat-y-var-est-part-1}
    \frac{1}{N^{2}}
    \sum_{i=1}^{n_{\text{P}}}\sum_{j=1}^{n_{\text{P}}}\frac{1}{\pi_{ij}}
    \widehat{\text{cov}}\left(
    \frac{1}{k}\sum_{t=1}^{k}Y_{\hat{\nu}(i, t)},
    \frac{1}{k}\sum_{t'=1}^{k}Y_{\hat{\nu}(j, t')}\right),
\end{equation}

\noindent assuming we have an estimate for:

\begin{equation*}
    \text{cov}\left(
    \frac{1}{k}\sum_{t=1}^{k}Y_{\hat{\nu}(i, t)},
    \frac{1}{k}\sum_{t'=1}^{k}Y_{\hat{\nu}(j, t')}\right).
\end{equation*}

The remaining term in \eqref{Law-total-variance-1-yhat-y} on the other hand is:
\begin{align*}
    &\frac{1}{N^{2}k^{2}}
    \mathbb{E}\left[\mathbb{V}\left[\left.
    \sum_{i \in S_{\text{P}}}\frac{1}{\pi_{i}}\sum_{t=1}^{k}Y_{\hat{\nu}(i, t)}\right|
    \sigma\left(\{Y_{i}, \bX_{i}, \hat{\nu}(i, t), \pi_{i}\}_{i\in U, t\in\{1,\dots,k\}}\right)\right]\right]=\\
    &\frac{1}{N^{2}k^{2}}
    \mathbb{E}\left[\mathbb{V}\left[\left.
    \sum_{i=1}^{N}\1_{\{i\in S_{\text{P}}\}}
    \frac{1}{\pi_{i}}\sum_{t=1}^{k}Y_{\hat{\nu}(i, t)}\right|
    \sigma\left(\{Y_{i}, \bX_{i}, \hat{\nu}(i, t), \pi_{i}\}_{i\in U, t\in\{1,\dots,k\}}\right)\right]\right]=\\
    &\frac{1}{N^{2}k^{2}}
    \sum_{i=1}^{N}
    \mathbb{E}\left[\frac{1}{\pi_{i}^{2}}\left(\sum_{t=1}^{k}Y_{\hat{\nu}(i, t)}\right)^{2}
    \mathbb{V}\left[\left.\1_{\{i\in S_{\text{P}}\}}\right|
    \sigma\left(\{Y_{i}, \bX_{i}, \hat{\nu}(i, t), \pi_{i}\}_{i\in U, t\in\{1,\dots,k\}}\right)\right]\right]+\\
    &\frac{1}{N^{2}k^{2}}
    \sum_{\substack{i,j=1\\i\neq j}}^{N}
    \mathbb{E}\left[\frac{1}{\pi_{i}\pi_{j}}\sum_{t',t=1}^{k}Y_{\hat{\nu}(i, t)}Y_{\hat{\nu}(j, t')}
    \text{cov}\left[\left.\1_{\{i\in S_{\text{P}}\}},\1_{\{j\in S_{\text{P}}\}}\right|
    \sigma\left(\{Y_{i}, \hat{\nu}(i, t), \bX_{i}, \pi_{i}\}_{i\in U, t\in\{1,\dots,k\}}\right)\right]\right]=\\
    &\frac{1}{N^{2}k^{2}}
    \sum_{i=1}^{N}
    \mathbb{E}\left[
    \frac{1-\pi_{i}}{\pi_{i}}
    \left(\sum_{t=1}^{k}Y_{\hat{\nu}(i, t)}\right)^{2}\right]+\\
    &\frac{1}{N^{2}k^{2}}
    \sum_{\substack{i,j=1\\i\neq j}}^{N}
    \mathbb{E}\left[\underbrace{\text{cov}\left(\1_{\{i\in S_{\text{P}}\}},\1_{\{j\in S_{\text{P}}\}}|\sigma(\pi_{i}, \bX_{i})_{i\in U}\right)}_{=\pi_{ij}-\pi_{i}\pi_{j}}\frac{1}{\pi_{i}\pi_{j}}
    \sum_{t',t=1}^{k}Y_{\hat{\nu}(i, t)}Y_{\hat{\nu}(j, t')}\right],
\end{align*}

\noindent which (if second order inclusion probabilities are known) can be estimated by:

\begin{align*}
    \frac{1}{N^{2}}\sum_{i \in S_{\text{P}}}\left(1-\pi_i\right)
    \frac{\displaystyle
    \left(\frac{1}{k}\sum_{t=1}^{k}y_{\hat{\nu}(i, t)}\right)^{2}}{\pi_i^2}
    +\frac{1}{N^{2}}\sum_{i \in S_{\text{P}}} \sum_{\substack{j \in S_{\text{P}} \\ j \neq i}}
    \frac{\pi_{ij}-\pi_i \pi_j}{\pi_{ij}}
    \frac{\displaystyle\frac{1}{k}\sum_{t=1}^{k}y_{\hat{\nu}(i, t)}}{\pi_i}
    \frac{\displaystyle\frac{1}{k}\sum_{t'=1}^{k}y_{\hat{\nu}(j, t')}}{\pi_j}.
\end{align*}
\end{proof}

\subsection{Proof of Theorem \ref{variance-pmm-proof-est}}\label{proof-of-3}

\begin{proof}
Conditionally on $\mathcal{G}$, the values $z_i$ are fixed and $\hat \mu = N^{-1} \sum_{i \in S_P} z_i/\pi_i$ is the HT estimator of $N^{-1} \sum_{1}^N z_i$. Notice that then the $\hat V_1$ is the HT variance estimator of $\hat\mu$.

Since $\mathbb{E}[\1_{\{i \in S_P\}}\1_{\{j \in S_P\}} \mid \mathcal{G}] = \pi_{ij}$ (where $\pi_{ii} = \pi_i$), we have:
\begin{align*}
    \mathbb{E}[\hat V_1\mid \mathcal{G}] &= \frac{1}{N^2} \sum_{i \in U} \sum_{j \in U} \frac{\pi_{ij} - \pi_i \pi_j}{\pi_{ij}} \frac{z_i z_j}{\pi_i \pi_j} \mathbb{E}[\1_{\{i \in S_P\}}\1_{\{j \in S_P\}} \mid \mathcal{G}] \\
    & = \frac{1}{N^2} \sum_{i \in U} \sum_{j \in U} \frac{\pi_{ij} - \pi_i \pi_j}{\pi_i \pi_j} z_i z_j = \mathbb{V}[\hat \mu \mid \mathcal{G}].
\end{align*}
By taking the expectation on both sides we get: $\mathbb{E}[\hat V_1] = V_1$ (recall the decomposition in Theorem \ref{variance-pmm-proof}). 

By assumption $\hat V_1 / \mathbb{V}[\hat \mu\mid \mathcal G] \overset{P}{\rightarrow} 1$ conditionally on $\mathcal{G}$. Set $\varepsilon > 0$ and denote $g_N = \mathbb{P}(|\hat V_1 / \mathbb{V}[\hat \mu \mid \mathcal G] - 1| > \varepsilon \mid \mathcal G)$. Notice that $0 \leq g_N \leq 1$ and $g_N \rightarrow 0$. Using the dominated convergence theorem, we get $\mathbb{P}(|\hat V_1 / \mathbb{V}[\hat \mu \mid \mathcal G] - 1| > \varepsilon) = \mathbb{E}[g_N] \rightarrow 0$ and therefore $\hat V_1 / \mathbb{V}[\hat \mu\mid \mathcal G] \overset{P}{\rightarrow} 1$ unconditionally.

Let $\theta_N = n_P \mathbb{V}[\hat\mu \mid \mathcal{G}]$. As per assumption $\left|(\pi_{ij}-\pi_{i}\pi_{j})\pi_{ij}^{-1}\right|$ are bounded and $\mathbb{E}[Y^4] < \infty$, the sequence $\{\theta_N\}$ is uniformly integrable. Then since also $\theta_N \overset{P}{\rightarrow} \gamma$ by assumption, we have $n_P V_1 = \mathbb{E}[\theta_N] \rightarrow \gamma$. Therefore:
\begin{equation*}
    \frac{\hat V_1}{V_1} = \underbrace{\frac{\hat V_1}{\mathbb{V}[\hat\mu \mid \mathcal{G}]}}_{\overset{P}{\rightarrow} 1} \underbrace{\frac{\theta_N}{n_PV_1}}_{\overset{P}{\rightarrow} \frac{\gamma}{\gamma} = 1} \overset{P}{\rightarrow} 1.
\end{equation*}

\end{proof}

\newpage

\section{Additional simulations}\label{app-additional-simulations}

In all simulations, for the PMM and the NN estimators we estimated the $V_2$
component using the proposed mini-bootstrap. The simulations in Sections~\ref{sec-appen-sim-3} and~\ref{sec-appen-sim-5} should be treated as exploratory, because not all of the results presented in the main paper have a~justification for these two scenarios.

In all simulation studies probability sample was selected using simple random sampling without replacement.

\subsection{Simple procedure for choosing \texorpdfstring{$k$}{k} hyper-parameter}\label{sec-appen-sim-1}

In this simulation study we provide evidence for choosing the $k$ hyper-parameter. We consider the following study design:

\begin{itemize}
    \item Population size: $N=100,000$,
    \item Probability sample size: $n_{\text{P}}=500$,
    \item Expected non-probability sample size: $\mathbb{E}\left[n_{\text{NP}}/N\right]\approx0.20$,
    \item $X_1$ was generated from the multivariate normal distribution transformed to 1d vector (the reason for that was to introduce correlation between observations):

     $\begin{pmatrix}
        \boldsymbol{Z}_{1}^{T} & \boldsymbol{Z}_{2}^{T} & \dots & \boldsymbol{Z}_{N/5}^{T}
    \end{pmatrix}
    $ where $\boldsymbol{Z}_{k}\sim\mathcal{N}_{5}\left(\begin{pmatrix}
        1 & 1 & 1 & 1 & 1
    \end{pmatrix}^{T},\boldsymbol{\Sigma}\right)$ ($\boldsymbol{Z}_{k}$ being i.i.d) with
    $$
        \boldsymbol{\Sigma}=\begin{pmatrix}
            1 & U_{12} & U_{13} & U_{14} & U_{15}\\
            U_{12} & 1 & U_{23} & U_{24} & U_{25}\\
            U_{13} & U_{23} & 1 & U_{34} & U_{35}\\
            U_{14} & U_{24} & U_{34} & 1 & U_{45}\\
            U_{15} & U_{25} & U_{35} & U_{45} & 1\\
        \end{pmatrix}
    $$ in setting where $U_{12},\dotso\sim\mathcal{U}\left(-\tfrac{1}{2},\tfrac{1}{2}\right)$ i.i.d.,
    \item $X_2 \sim\text{Exp}(1)$ i.i.d,
    \item Vector of $\epsilon$'s analogously to $x_{1}$ being $\begin{pmatrix}
        \boldsymbol{R}_{1}^{T} & \boldsymbol{R}_{2}^{T} & \dots & \boldsymbol{R}_{N/5}^{T}
    \end{pmatrix}^{T}$ where $\boldsymbol{R}_{k}\sim\mathcal{N}_{5}\left(\boldsymbol{0},\boldsymbol{\Sigma}\right)$ and  $U_{12},\dotso\sim\mathcal{U}\left(-0.7,1\right)$ i.i.d.,
    \item We generated two vector of probabilities for selection into the non-probability sample $p_{1}=\dfrac{\exp(x_{2})}{1+\exp(x_{2})}$ and $p_{2}=\dfrac{\exp(x_{1})}{1+\exp(x_{1})}$,
    \item The inclusion into non-probability sample $S_{\text{NP}}$ was generated as 

    $$
    \delta=\min\left\{\text{Bernoulli}(p_{1}), \text{Bernoulli}(p_{2}),\1_{\{\epsilon>\epsilon_{.8}\}}+\1_{\{\epsilon<\epsilon_{.2}\}}\right\} 
    $$
    where $\epsilon_{t}$ is an empirical quantile of $\epsilon$ of order $t$.
    \item Two target variables were generated according to the following formulas:
    \begin{align*}
        Y_1 & = 1+\tfrac{1}{2}x_{1}+\tfrac{35}{100}x_{2}+\epsilon\\
        Y_2 & = 1.2+(x_{1}-0.5)^{2}+\arctan(x_{2})^{3+\sin(x_{1}+x_{2})}+
    \sin(x_{1})\cos(x_{2})+\epsilon,
    \end{align*}
\end{itemize}

\noindent and $k$ being chosen by minimising $\widehat{\mathbb{V}\left[\hat{\mu}\right]}$ in each simulation iteration. We consider four estimators discussed in the paper: GLM, NN, PMM A ($\hat{y}-\hat{y}$ matching) and PMM B ($\hat{y}-y$ matching). The NN and PMM estimators are with fixed $k=5$ and dynamic selection of $k$. 

Table \ref{tab-appendix-sim1} shows the results of the simulation study. As regards the first variable ($Y_1$), slight differences are observed between the estimators under consideration. The dynamic selection is characterised by a smaller bias for the NN estimator and a slightly better RMSE for the NN and PMM A estimators, compared to the scenario with fixed $k$. Average $k$ for the NN estimator is large, about 27, while for the PMM A it is around 23 and for the PMM B around~1.5. For this linear variable, however, the lowest RMSE is achieved by the GLM and PMM B estimators.

\begin{table}[ht]
\centering
\caption{Results for the simulation study on choosing of $k$ hyper-parameter (all measures but $\bar{k}$ are multiplied by 100)} 
\label{tab-appendix-sim1}
\begin{tabular}{lllrrrrr}
  \hline
Variable & Estimator & $k$-selection & Bias & SE & RMSE & CR & $\bar{k}$ \\ 
  \hline
Y1 & GLM & -- & -0.9163 & 4.0726 & 4.1744 & 97.80 & -- \\ 
   & NN & Dynamic & 1.0887 & 4.4024 & 4.5351 & 86.20 & 26.66 \\ 
   &  & Fixed & 1.1371 & 5.7830 & 5.8937 & 91.40 & -- \\ 
   & PMM A & Dynamic & 0.9618 & 4.5397 & 4.6404 & 86.60 & 23.20 \\ 
   &  & Fixed & 0.7895 & 5.7732 & 5.8269 & 91.00 & -- \\ 
   & PMM B & Dynamic & -0.9268 & 4.0754 & 4.1794 & 85.40 & 1.54 \\ 
   &  & Fixed & -0.9260 & 4.0740 & 4.1779 & 85.60 & -- \\ 
  Y2 & GLM & -- & -22.1792 & 8.2605 & 23.6676 & 26.80 & -- \\ 
   & NN & Dynamic & -5.5973 & 8.2556 & 9.9742 & 84.20 & 18.92 \\ 
   &  & Fixed & -2.3018 & 9.0721 & 9.3595 & 92.80 & -- \\ 
   & PMM A & Dynamic & -2.4117 & 8.1857 & 8.5336 & 93.00 & 14.47 \\ 
   &  & Fixed & -1.5832 & 8.5206 & 8.6665 & 94.20 & -- \\ 
   & PMM B & Dynamic & -22.1395 & 8.2436 & 23.6245 & 19.60 & 4.17 \\ 
   &  & Fixed & -22.1554 & 8.2521 & 23.6423 & 19.60 & -- \\ 
   \hline
\end{tabular}
\end{table}
In the case of the highly non-linear variable $Y_2$, both GLM and PMM B estimators are significantly biased, while the NN and PMM A, with either dynamic or fixed $k$, are characterised by negligible bias. In addition, PMM A with dynamic selection of $k$ is characterised by a~smaller RMSE. The average $k$ values for $Y_2$ are somewhat lower for the NN and PMM A estimators than for $Y_1$.

\clearpage

\subsection{Simulation with variable selection}\label{sec-appen-sim-2}

In this case our study design follows that proposed by \citet{yang_doubly_2020}:

\begin{itemize}
    \item finite population size was set to $N=10000$,
    \item for each unit we generate $X_{i}=\begin{pmatrix}
        1 & X_{1, i} & \ldots & X_{p-1, i}
    \end{pmatrix}^{T}$ where $p=50$ and $X_{i,t}\sim\mathcal{N}(0,1)$ i.i.d for each $i$ and $t=0,\dots,p-1$,
    \item from the finite population, we select a non-probability sample $S_{\text{NP}}$ of size $n_{\text{NP}} \approx 2000$, according to the selection indicator $I_{\text{NP}, i} \sim$  $\operatorname{Bernoulli}\left(\pi_{\text{NP}, i}\right)$,
    \item we select a probability sample $S_{\text{P}}$ of the average size $n_{\text{P}}=500$ under Poisson sampling with $\pi_{\text{P}, i} \propto\left(0.25+\left|X_{1 i}\right|+0.03\left|Y_{i}\right|\right)$ (we have 4 sets of $\pi_{\text{P}}$ as we have 4 $Y$ variables defined below),
    \item for the non-probability sample inclusion probability, we consider both linear and non-linear sampling score models:

    \begin{itemize}
        \item  $\operatorname{logit}\left(\pi_{\text{NP}, i}\right)=\balpha_{0}^{\mathrm{T}} \bX_{i}$, where $\balpha_{0}=\begin{pmatrix}
            -2 & 1 & 1 & 1 & 1 & 0 & 0 & \ldots &  0
        \end{pmatrix}^{\mathrm{T}}$ (model PSM I)
        \item $\operatorname{logit}\left(\pi_{\text{NP}, i}\right)=3.5+\balpha_{0}^{T} \log \left(\bX_{i}^{2}\right)-\sin \left(X_{3, i}+X_{4, i}\right)-X_{5, i}-X_{6, i}$,\\ where $\balpha_{0}=
        \begin{pmatrix}
            0 & 0 & 0 & 3 & 3 & 3 & 3 & 0 & \ldots & 0
        \end{pmatrix}^{T}$  (model PSM II)
    \end{itemize}

    \item to generate a continuous outcome variable $Y_{i}$, we consider both linear and non-linear outcome models with $\beta_{0}=\begin{pmatrix}
        1 & 0 & 0 & 1 & 1 & 1 & 1 & 0 & \ldots &  0
    \end{pmatrix}^{T}$ 

    \begin{itemize}
        \item $Y_{i}=\bbeta_{0}^{\mathrm{T}} \bX_{i}+\epsilon_{i}, \epsilon_{i} \sim \mathcal{N}(0,1)$ i.i.d (model OM I, denoted as $Y_{11}$),
        \item $Y_{i}=1+\exp \left(3 \sin \left(\bbeta_{0}^{T} \bX_{i}\right)\right)+X_{5, i}+X_{6, i}+\epsilon_{i}, \epsilon_{i} \sim \mathcal{N}(0,1)$ i.i.d (model OM II, denoted as $Y_{12}$),
    \end{itemize}
    \item to generate a binary outcome variable $Y_{i}$, we consider both linear and non-linear outcome models with $\bbeta_{0}=\begin{pmatrix}
        1 & 0 & 0 & 3 & 3 & 3 & 3 & 0 & \ldots & 0
    \end{pmatrix}^{T}$:
    \begin{itemize}
        \item $Y_i \sim \operatorname{Bernoulli}\left(\pi_{Y}(\bX_i)\right)$ with logit $\left(\pi_{Y}(\bX_i)\right)=\bbeta_{0}^{T} \bX_i$ (model OM III, denoted as $Y_{21}$),
        \item  $Y_i \sim \operatorname{Bernoulli}\left(\pi_{Y}(\bX_i)\right)$ with logit $\left(\pi_{Y}(\bX_i)\right)=2-\log\left(\left(\bbeta_{0}^{T} \bX_i\right)^{2}\right)+2 X_{5, i}+2 X_{6, i}$ (model OM IV, denoted as $Y_{22}$).
    \end{itemize}
    \item variable selection was done using SCAD penalty.
\end{itemize}

Table \ref{tab-appendix-sim2} shows the results of the simulation study with variable selection. As expected, when the model is linear (OM I and OM III), GLM, PMM A and PMM B (except for OM III) are almost unbiased with coverage close to the nominal rate. The NN estimator without variable selection for the OM I and OM III scenarios is significantly biased and the bias decreases with the variable selection.  The PMM B ($\hat{y}-y$ matching) estimator is characterised by greater bias for both linear and non-linear selection mechanisms with comparable levels as the NN estimator.  Variable selection reduces the bias and slightly reduces the standard error for the estimators discussed. As a result, the RMSE for OM I and OM III is slightly lower than without variable selection.

\begin{table}[ht!]
\centering
\caption{Results of the simulation with variable selection (all numbers multiplied by 100)} 
\label{tab-appendix-sim2}
\begin{tabular}{llrrrrrrrr}
  \hline
   & & \multicolumn{4}{c}{Linear PS (PSM I)} & \multicolumn{4}{c}{Non-linear PS (PSM II)}\\
Estimator & Selection & Bias & SE & RMSE & CR & Bias & SE & RMSE & CR \\
 \hline
\multicolumn{10}{c}{$Y_{11}$ (OM I)} \\
 \hline
GLM & No & 0.66 & 10.29 & 10.31 & 95.40 & 0.23 & 10.08 & 10.08 & 95.80 \\ 
   & Yes & -0.16 & 10.13 & 10.13 & 96.00 & 0.38 & 9.95 & 9.96 & 96.40 \\ 
  NN & No & 64.18 & 8.19 & 64.70 & 0.00 & -36.98 & 9.06 & 38.07 & 1.60 \\ 
   & Yes & 64.18 & 8.19 & 64.70 & 0.00 & -36.98 & 9.06 & 38.07 & 1.60 \\ 
  PMM A & No & 1.35 & 10.70 & 10.79 & 93.40 & 0.54 & 10.48 & 10.50 & 94.20 \\ 
   & Yes & 0.33 & 10.38 & 10.38 & 94.80 & 0.26 & 10.17 & 10.18 & 95.20 \\ 
  PMM B & No & 0.87 & 10.27 & 10.31 & 94.40 & 0.22 & 10.08 & 10.08 & 95.20 \\ 
   & Yes & 0.06 & 10.11 & 10.11 & 95.40 & 0.38 & 9.95 & 9.96 & 96.00 \\ 
   \hline
\multicolumn{10}{c}{$Y_{12}$ (OM II)} \\
 \hline
GLM & No & 96.92 & 24.84 & 100.06 & 3.00 & -101.28 & 12.52 & 102.06 & 0.00 \\ 
   & Yes & 80.99 & 25.20 & 84.81 & 11.00 & -97.60 & 10.59 & 98.18 & 0.00 \\ 
  NN & No & 102.26 & 29.78 & 106.51 & 3.00 & -91.86 & 22.38 & 94.54 & 1.80 \\ 
   & Yes & 102.26 & 29.78 & 106.51 & 3.00 & -91.86 & 22.38 & 94.54 & 1.80 \\ 
  PMM A & No & 97.06 & 29.86 & 101.55 & 4.60 & -97.65 & 20.01 & 99.68 & 0.60 \\ 
   & Yes & 92.79 & 29.47 & 97.36 & 6.40 & -77.56 & 20.96 & 80.34 & 7.00 \\ 
  PMM B & No & 96.91 & 24.84 & 100.04 & 1.00 & -101.30 & 12.52 & 102.07 & 0.00 \\ 
   & Yes & 81.03 & 24.60 & 84.68 & 2.20 & -97.59 & 10.54 & 98.16 & 0.00 \\ 
   \hline
\multicolumn{10}{c}{$Y_{21}$ (OM III)} \\
 \hline
GLM & No & -0.79 & 2.15 & 2.30 & 95.00 & 1.10 & 1.73 & 2.06 & 92.40 \\ 
   & Yes & 0.24 & 1.88 & 1.89 & 96.60 & 0.98 & 1.68 & 1.94 & 94.00 \\ 
  NN & No & 10.18 & 2.02 & 10.37 & 0.20 & -7.34 & 1.73 & 7.54 & 3.00 \\ 
   & Yes & 10.18 & 2.02 & 10.37 & 0.20 & -7.34 & 1.73 & 7.54 & 3.00 \\ 
  PMM A & No & -0.84 & 2.39 & 2.53 & 94.60 & 1.04 & 1.91 & 2.17 & 98.80 \\ 
   & Yes & 0.27 & 2.20 & 2.22 & 97.40 & 0.96 & 1.88 & 2.11 & 99.20 \\ 
  PMM B & No & 2.75 & 3.03 & 4.10 & 86.20 & 4.82 & 2.50 & 5.43 & 67.00 \\ 
   & Yes & 4.55 & 2.84 & 5.36 & 68.20 & 5.41 & 2.55 & 5.98 & 59.40 \\ 
   \hline
\multicolumn{10}{c}{$Y_{22}$ (OM IV)} \\
 \hline
GLM & No & 1.07 & 1.97 & 2.24 & 94.00 & -6.52 & 1.67 & 6.73 & 4.60 \\ 
   & Yes & 1.73 & 1.82 & 2.51 & 90.80 & -6.32 & 1.60 & 6.52 & 5.20 \\ 
  NN & No & -1.28 & 2.06 & 2.42 & 91.20 & 0.26 & 1.59 & 1.61 & 100.00 \\ 
   & Yes & -1.28 & 2.06 & 2.42 & 91.20 & 0.26 & 1.59 & 1.61 & 100.00 \\ 
  PMM A & No & 0.98 & 2.15 & 2.36 & 98.80 & -7.38 & 1.86 & 7.62 & 10.60 \\ 
   & Yes & 1.69 & 2.02 & 2.63 & 96.20 & -7.49 & 1.88 & 7.72 & 8.60 \\ 
  PMM B & No & 3.70 & 2.56 & 4.50 & 82.60 & -4.08 & 2.56 & 4.82 & 77.00 \\ 
   & Yes & 4.73 & 2.42 & 5.31 & 71.80 & -3.19 & 2.52 & 4.06 & 85.20 \\ 
   \hline
\end{tabular}
\end{table}

In the case of non-linear OM II and OM IV, all the proposed estimators are characterised by large bias and variance. Furthermore, in this case variable selection leads to inconclusive results. For example, for the linear PSM I and OM II, variable selection reduces the bias, and for some models it also reduces the standard error. For the non-linear PSM II, bias and standard error are slightly reduced.  In the case of binary (OM IV) and linear selection (PSM I), the GLM and PMM A estimators have a coverage close to nominal, but variable selection increases bias while decreasing standard error. For this combination, the RMSE is larger for the estimators with variable selection. For non-linear selection (PSM II), all estimators are significantly biased and the coverage is close to zero, except for the NN and PMM B estimator. Interestingly, the CR for the NN estimator for the OM IV and PSM II is close to the nominal rate which requires further studies.

\clearpage
\subsection{Non-parametric regression methods}\label{sec-appen-sim-3}

In this section we check the performance of the PMM estimators with non-parametric regression through LOESS via \texttt{stats::loess} function in R with the following specification to boost performance: 
\begin{itemize}
    \item \texttt{loess.control(surface="interpolate")} -- surface approximation using a KD-tree ,
    \item \texttt{loess.control(trace.hat="interpolate")} -- approximated trace of the smoother matrix,
    \item span set to 0.2.
\end{itemize}

 We note that, formally, the theory developed in the main paper does not carry over to the non-parametric case. For PMM~B the situation is clearer, since transferring the assumptions and consistency theorems appears more straightforward and, in fact, feasible. For PMM~A, by contrast, the linear representation assumption~\eqref{linear-rep-assum} becomes problematic and would require a~substitute (potentially impossible to verify in practice), such as the leave-one-out stability condition discussed in the main paper. For this reason, this study should be regarded as an exploratory experiment rather than a formal confirmation of the theoretical properties of the estimators.
  
We adapt the same setting as in section \ref{sec-appen-sim-1} and define additional target variable $Y_{3}=x_{1}x_{2}\epsilon$ (with $\mathbb{E}(Y_3)=0$). Table \ref{tab-appen-sim-3} shows the results of this simulation study. As expected, the proposed PMM estimators based on non-parametric regression are characterised by negligible bias, but the variance is significantly larger than that of the GLM estimator for $Y_1$. The NN estimator is characterised with significantly larger variance than the proposed estimators.

\begin{table}[ht]
\centering
\caption{Results for the simulation study with non-parametric predictive mean matching models (all numbers multiplied by 100)} 
\label{tab-appen-sim-3}
\begin{tabular}{lllrrrr}
  \hline
Y & Estimator & Non-parametric? & Bias & SE & RMSE & CR \\ 
  \hline
Y1 & GLM & No & -0.95 & 4.07 & 4.18 & 96.80 \\ 
   & NN & Yes & 2.75 & 9.96 & 10.33 & 93.20 \\ 
   & PMM A & No & 0.64 & 5.79 & 5.82 & 92.20 \\ 
   &  & Yes & 0.59 & 5.68 & 5.71 & 92.20 \\ 
   & PMM B & No & -0.96 & 4.08 & 4.19 & 85.80 \\ 
   &  & Yes & 0.47 & 4.15 & 4.17 & 83.60 \\ 
  Y2 & GLM & No & -21.84 & 8.21 & 23.33 & 26.00 \\ 
   & NN & Yes & 1.36 & 12.44 & 12.52 & 94.40 \\ 
   & PMM A & No & -0.62 & 9.32 & 9.34 & 92.20 \\ 
   &  & Yes & 0.17 & 9.90 & 9.90 & 92.80 \\ 
   & PMM B & No & -21.81 & 8.20 & 23.30 & 18.20 \\ 
   &  & Yes & 1.85 & 8.67 & 8.86 & 94.20 \\ 
  Y3 & GLM & No & 1.95 & 3.23 & 3.77 & 99.40 \\ 
   & NN & Yes & 7.22 & 17.94 & 19.34 & 93.20 \\ 
   & PMM A & No & 7.33 & 9.33 & 11.87 & 88.20 \\ 
   &  & Yes & 4.37 & 9.68 & 10.62 & 89.80 \\ 
   & PMM B & No & 1.95 & 3.23 & 3.77 & 62.80 \\ 
   &  & Yes & 4.57 & 4.04 & 6.10 & 42.80 \\ 
   \hline
\end{tabular}
\end{table}

As expected, the bias and variance for GLM are large compared to the non-parametric PMM estimators for the non-linear variable $Y_2$. Note that the PMM A ($\hat{y}-\hat{y}$ matching) estimator is almost unbiased even when the target variables are highly non-linear. For the last variable $Y_3$ (with $\mathbb{E}(Y_3)=0$), the GLM estimator over-covers, whereas the PMM~B estimator markedly under-covers.

\clearpage
\subsection{Violation of the positivity assumption}\label{sec-appen-sim-4}

In this section we check the performance of the proposed estimators when the positivity assumption is violated. We consider the same settings as in section \ref{sec-appen-sim-1} with additional $X_3$ variable and effects connected with this variable. 

\begin{itemize}
    \item $X_3 \sim \text{NegBin}(\mu=10, r=4)$ and then numbers over 40 were grouped into one category "40",
    \item we created two target variables:
    \begin{itemize}
        \item $Y_1 = -7+6X_{1}-5X_{2}+X_{3}^{T}\gamma+15\epsilon$,
        \item $Y_{2} = -2+0.37\cdot (X_{1}-0.5)^{2}+X_{2}^{2}+X_{3}^{T}\gamma+5\epsilon$,
    \end{itemize}
    where $\gamma \sim \mathcal{U}\left(-6, 10\right)$ are coefficients associated with each level of $X_3$ variable (40 possible levels),
    \item we consider two scenarios for under-coverage:
    \begin{itemize}
        \item \textit{Stochastic}: where the samplable sub-population is generated from $\text{Binomial}(\pi_{u})$ where $\logit(\pi_{u})\allowbreak=X_3^T\lambda$ and $\lambda = \begin{pmatrix}
            0.75 & 0.72 & \ldots & 0.10
        \end{pmatrix}^{T}$, so the larger the $X_3$ the less likely to be observed in the sub-population.
        \item \textit{Deterministic}: where the samplable sub-population is generated in the following process:
        \begin{itemize}
            \item target variable $\pi_{u}$ is defined as $\logit(\pi_{u})=0.84 + 0.32 X_1 + 0.68 X_2 + X_3^T\lambda$ where $\lambda$ is defined above,
            \item inclusion indicator was generated as $\1_{\{\pi_{u} > Q_{\pi_{u}, 0.25}\}}$, where $Q_{\pi_{u}, 0.25}$ is the 25\% percentile of $\pi_{u}$.
        \end{itemize}
    \end{itemize}
    \item probabilities of inclusion in the non-probability sample from the under-represented populations were defined as follows:
    \begin{itemize}
        \item \textit{Stochastic}:  $\logit(p_i)= X_{2}-X_{1}-2$, so the expected size is about 17\% of the population.
        \item \textit{Deterministic}: $\logit(p_i) = 0.6 X_{1}-X_{2}-2$, so the expected size is about 20\% of the population.
    \end{itemize}
\end{itemize}

Table \ref{tab-appen-sim-4} shows the results of the simulation study when the positivity assumption is violated. For the linear $Y_1$ all the estimators yield the same results with slightly lower bias for the PMM B ($\hat{y}-y$ fit). However, the PMM A suffers from larger standard errors and, as a result, the standard error and RMSE are larger for deterministic and stochastic results and the CR is above the nominal level (about 92\%). For the non-linear $Y_2$, the main difference exists for deterministic under-coverage. The bias of PMM A is lower than that of the MI-GLM and PMM B estimators and the standard error is comparable. Under this scenario (PMM A), coverage is the highest (about 58\%), but still does not reach the nominal level of 95\%. In the case of stochastic under-coverage, the proposed estimators are comparable to the MI-GLM estimator, with a slightly smaller bias for the PMM A estimator.

\begin{table}[ht!]
\centering
\caption{Results of the simulation study when the positivity assumption is violated (results multiplied by 100)} 
\label{tab-appen-sim-4}
\begin{tabular}{lllrrrr}
  \hline
Variable & Under-coverage & Estimator & Bias & SE & RMSE & CR \\ 
  \hline
Y1 & Deterministic & GLM & -13.92 & 49.62 & 51.53 & 97.00 \\ 
   &  & NN & -15.88 & 70.64 & 72.40 & 90.20 \\ 
   &  & PMM A & -10.02 & 68.74 & 69.46 & 92.20 \\ 
   &  & PMM B & -13.91 & 49.62 & 51.53 & 90.00 \\ 
   & Stochastic & GLM & 7.75 & 51.07 & 51.66 & 97.00 \\ 
   &  & NN & 16.68 & 68.51 & 70.52 & 89.40 \\ 
   &  & PMM A & 7.86 & 68.84 & 69.29 & 91.60 \\ 
   &  & PMM B & 7.75 & 51.08 & 51.66 & 88.60 \\ 
  Y2 & Deterministic & GLM & -82.73 & 31.42 & 88.49 & 30.40 \\ 
   &  & NN & -0.36 & 32.72 & 32.72 & 91.80 \\ 
   &  & PMM A & -53.32 & 32.64 & 62.52 & 57.60 \\ 
   &  & PMM B & -82.73 & 31.42 & 88.49 & 24.80 \\ 
   & Stochastic & GLM & 13.12 & 28.35 & 31.24 & 95.80 \\ 
   &  & NN & -7.00 & 31.31 & 32.09 & 94.20 \\ 
   &  & PMM A & 11.42 & 30.95 & 32.99 & 93.80 \\ 
   &  & PMM B & 13.11 & 28.34 & 31.23 & 92.40 \\ 
   \hline
\end{tabular}
\end{table}

The deterministic case is supposed to mirror online sampling (or sampling via telephone connection) where a unit can only be sampled if it has the internet connection (a telephone number). Such cases usually depend on age, mirrored by $X_{3}$, and, to a smaller extent, on social status, which may be correlated with covariates. Note that despite the violation of the positivity assumption in the deterministic case $\mathbb{P}\left[\left.\delta=1\right|\bX, Y\right]= \mathbb{P}\left[\left.\delta=1\right|\bX\right]$ a.s is maintained. In the case of stochastic under-coverage, the selection of units to the non-probability sample can be viewed as a two-stage sampling process; in this case, positivity is not violated, which explains good results of the MI-GLM estimator.

\clearpage

\subsection{Multiply robust imputation}\label{sec-appen-sim-5}

In this section, we test whether the approach proposed by \citet{chen_note_2021} is suitable for non-probability samples. They proposed multiply robust imputation using predictive mean matching, where predictions from multiple models are used as independent variables. In this section, we follow the same design as that described in section \ref{sec-appen-sim-1}, but we create two target variables as follows:

\begin{itemize}
    \item $Y_{1}=1+\tfrac{1}{5}X_{1}+5X_{2}+\epsilon$,
    \item $Y_{2}=-2+5\left(X_{1}-0.5\right)^{5}+X_{2}^{3}+\epsilon$.
\end{itemize}

For both models we consider two regression models: 

\begin{itemize}
    \item Model 1: linear regression with $X_1$ and $X_2$,
    \item Model 2: linear regression with $(X_1-0.5)^5$ and $X_2^3$,
\end{itemize}

\noindent so that for each model, one of the models considered is correctly specified. For both PMM estimators without the multiple robust property, we used a linear model with $x_1$ and $x_2$.

\begin{table}[ht!]
\centering
\caption{Results for the simulation study with multiply robust predictive mean matching} 
\label{tab-appen-sim-5}
\begin{tabular}{lllrrrr}
  \hline
Variable & Estimator & MultiRobust & Bias & SE & RMSE & CR \\ 
  \hline
$Y_1$ & GLM & -- & -0.01 & 0.23 & 0.23 & 93.20 \\ 
   & NN & -- & 0.00 & 0.22 & 0.22 & 94.00 \\ 
   & PMM A & No & -0.01 & 0.23 & 0.23 & 93.40 \\ 
   &  & Yes & -0.01 & 0.23 & 0.23 & 93.40 \\ 
   & PMM B & No & -0.01 & 0.23 & 0.23 & 93.20 \\ 
   &  & Yes & -0.01 & 0.23 & 0.23 & 93.20 \\ 
  $Y_2$ & GLM & -- & -15.38 & 5.76 & 16.43 & 28.20 \\ 
   & NN & -- & 0.06 & 8.03 & 8.03 & 94.20 \\ 
   & PMM A & No & -0.47 & 7.90 & 7.91 & 92.80 \\ 
   &  & Yes & 0.51 & 8.42 & 8.44 & 94.20 \\ 
   & PMM B & No & -14.17 & 5.68 & 15.27 & 32.80 \\ 
   &  & Yes & 0.52 & 8.42 & 8.44 & 94.00 \\ 
   \hline
\end{tabular}
\end{table}

Table \ref{tab-appen-sim-5} contains the results for multiply robust predictive mean matching. All linear estimators are characterised by the same bias, SE, RMSE and CR. Differences can be observed for the non-linear $Y_2$. GLM and the standard PMM B are characterised by similar bias and RMSE, while the addition of multiplicative robustness to the PMM B estimator significantly reduces its bias. For the PMM A estimator, it does not matter whether the fitting is done with a linear model with $x_1$ and $x_2$ or with predictions from the two models considered. In this limited simulation study, none of the models for $Y_2$ reached the nominal coverage rate of 95\%. For $Y_2$, the NN estimator is characterised by one of the lowest RMSE values and the highest CR.

\clearpage
\section{Implementation in the R language}

The proposed methods are implemented in the R language in the \texttt{nonprobsvy} package. The 0.3.0 version is available on CRAN but in this paper we used the development version. One should install the package from the github repository using the following code:

\begin{verbatim}
install.packages("pak")
pak::install_github("ncn-foreigners/nonprobsvy") ## 0.3.0
\end{verbatim}

Below we present functions for the PMM estimator under specific settings. In general, the user should change the following arguments:

\begin{itemize}
    \item \texttt{method\_outcome = "pmm"} -- specification that the PMM estimator is used
    \item \texttt{control\_out} -- controls for MI estimator
    \begin{itemize}
        \item \texttt{k} -- number of nearest neighbours,
        \item \texttt{predictive\_match} -- whether $\hat{y}-\hat{y}$ (=1; PMM A) or $\hat{y}-y$ (=2; PMM B) matching should be used,
        \item \texttt{pmm\_reg\_engine} -- \texttt{"glm"} or \texttt{"loess"},
        \item \texttt{pmm\_k\_choice} -- automated selection of $k$ neighbours; use \texttt{"none"} or \texttt{"min\_var"},
        \item \texttt{pmm\_k\_max} -- upper bound for the \texttt{"min\_var"} argument.
    \end{itemize}
    \item \texttt{control\_inf} -- controls for inference (variance estimation)
    \begin{itemize}
        \item \texttt{pmm\_exact\_se} -- whether $\hat{V}_2$ should be estimated using "mini" bootstrap (=\texttt{TRUE}) or not (=\texttt{FALSE}).
    \end{itemize}
\end{itemize}

\clearpage

Example codes are shown below.

\begin{itemize}
    \item Predictive mean matching with $\hat{y}-\hat{y}$ matching (PMM A) with $\hat{V}_2$ estimated using "mini"-bootstrap presented in Section \ref{sec-main}.
    \begin{verbatim}
PMM_A <- nonprob(
  outcome = y1 ~ x1 + x2,
  data = sample_nonprob,
  svydesign = sample_prob,
  method_outcome = "pmm",
  family_outcome = "gaussian",
  control_outcome = control_out(k = 5, 
                               predictive_match = 1),
  control_inference = control_inf(pmm_exact_se = TRUE)
)
\end{verbatim}

\item Predictive mean matching with $\hat{y}-y$ matching (PMM B) with $\hat{V}_2$ estimated using "mini"-bootstrap presented in Section \ref{sec-main}.

\begin{verbatim}
PMM_B <- nonprob(
  outcome = y1 ~ x1 + x2,
  data = sample_nonprob,
  svydesign = sample_prob,
  method_outcome = "pmm",
  family_outcome = "gaussian",
  control_outcome = control_out(k = 5, 
                               predictive_match = 2),
  control_inference = control_inf(pmm_exact_se = TRUE)
)
\end{verbatim}

\item Predictive mean matching with $\hat{y}-\hat{y}$ matching with local linear regression (PMM A) with $\hat{V}_2$ estimated using "mini"-bootstrap presented in Section \ref{sec-main}.
 
\begin{verbatim}
PMM_A <- nonprob(
  outcome = y1 ~ x1 + x2,
  data = sample_nonprob,
  svydesign = sample_prob,
  method_outcome = "pmm",
  family_outcome = "gaussian",
  control_outcome = control_out(k = 5, 
                               predictive_match = 1,
                               pmm_reg_engine = "loess"),
  control_inference = control_inf(pmm_exact_se = TRUE)
)
\end{verbatim}

Furthermore, we would like to acknowledge the authors of the following packages that were utilized in this paper: \texttt{data.table}, \texttt{sampling}, \texttt{doSNOW}, \texttt{progress}, \texttt{foreach}, and \texttt{xtable}. 
\end{itemize}

\end{document}